# Performance-related differences of bibliometric statistical properties of research groups: cumulative advantages and hierarchically layered networks

Anthony F. J. van Raan
Centre for Science and Technology Studies
Leiden University
Wassenaarseweg 52
P.O. Box 9555
2300 RB Leiden, The Netherlands

***Abstract***

In this paper we distinguish between top-performance and lower performance groups in the analysis of statistical properties of bibliometric characteristics of two large sets of research groups. We find intriguing differences between top-performance and lower performance groups, but also between the two sets of research groups. Particularly these latter differences are interesting, as they may indicate the influence of research management strategies. Lower performance groups have a larger scale-dependent cumulative advantage than top-performance groups. We also find that regardless of performance, larger groups have less not-cited publications. We introduce a simple model in which processes at the micro level lead to the observed phenomena at the macro level. Top-performance groups are, on average, more successful in the entire range of journal impact. We fit our findings into a concept of hierarchically layered networks. In this concept, the network of research groups constitutes a layer of one hierarchical step higher than the basic network of publications connected by citations. The cumulative size-advantage of citations received by a group looks like preferential attachment in the basic network in which highly connected nodes (publications) increase their connectivity faster than less connected nodes. But in our study it is size that causes an advantage. In general, the larger a group (node in the research group network), the more incoming links this group acquires in a non-linear, cumulative way. Moreover, top-performance groups are about an order of magnitude more efficient in creating linkages (i.e., receiving citations) than the lower performance groups.

## 1. Introduction

In a recent paper (van Raan 2005a) we presented an empirical approach to the study of the statistical properties of bibliometric indicators on a very relevant but not simply 'available' aggregation level: the research group. Research groups are defined by the internal structure of universities, research institutions, R&D laboratories of companies, and thus they are not an entity directly available in databases such as authors or journals. The focus of our earlier study was on the distribution functions of a coherent set of indicators frequently used as a measuring instrument in the analysis of research performance, in order to provide a better insight into the statistical properties of the instrument.

Starting with the most basic statistical element in bibliometric analysis, the very skew distribution of citations over publications, we clearly observed in our earlier study the working of the central limit theorem. We found that at the level of research groups the distribution functions of the main indicators, particularly the journal-normalized and the field-normalized indicators, are approaching normal distributions. The results underlined the importance of the idea of 'group oeuvre', i.e., the role of sets of *organizationally* related publications as a unit of analysis. We notice that organizationally related publications differ from co-author related or citation related publications. These latter types of relations are the basis of almost all publication-data based networks in current network studies. The difference with organizationally related publications is precisely the reason why we elaborate the idea of *hierarchically layered networks*. We come back to this concept in Section 3.3,

In this follow-up study we take a more differentiated approach. Our aim is to relate statistical properties of bibliometric indicators with different aspects of the scientific communication system. Central questions are to what extent the empirically found distribution and correlation functions are reflections of science-internal properties of the reward system, for instance, how scientific work is valued by colleagues; and of how scientists disseminate their findings as optimal as possible, more specifically, how these flows 'must' go because of general network properties. But these distribution and correlation functions may also reflect science-external properties, such as the consequences of specific research management strategies.

An even more fundamental question is 'which functions do bibliometric entities have?' For instance, the role of references, and with that, the meaning of citations, is probably a more-dimensional characteristic, ranging from direct recognition to symbols for specific research themes, and from utility to persuasion. Scientists refer less to 'authoritative' papers in the case of smaller total numbers of references (Moed and Garfield 2004). From this empirical finding follows that citing such authoritative papers is not a major motivation of an author. As these authoritative papers are often highly cited, the above phenomenon may cause a mitigation of the accumulation of citations to already highly cited papers.

Katz (1999, 2000, 2005) discusses scaling relationships between number of citations and number of publications across research fields, institutes and countries. He concludes that the scientific community is characterized by the 'Matthew effect' (cumulative advantage, Merton 1968, 1988) implying a non-linear increase of impact with increasing size, demonstrated by the finding that the number of citations as a function of number of publications (measured for 152 fields of science) exhibits a power law dependence with an exponent larger than 1. In our earlier paper (van Raan 2005a) we showed that *also at the level of research groups* a size-dependent cumulative advantage of the correlation between number of citations and number of publications exists.

The next step, presented in this paper, is to investigate these size-dependent effects as a function of crucial parameters such as research performance. From a general viewpoint it is important to study size-dependent characteristics of a system that is basically described by fractal (power-law distribution) properties. More specifically, often the scientific communication system is described as a network of publications with a fractal topology (in terms of the in-degree distributions of the links between the publications as nodes in the network), and thus the system is considered to be ‘scale-free’ (Zitt *et al* 2003, 2004; Zitt 2005). Nevertheless, our observations indicate that size (and thus scale-) dependent characteristics do exist in such a system, be it in the case specific clusters of publications, namely research groups.

An interesting question in this context is whether indicators that are more complex than simple citation counts -such as our field-specific normalized performance indicator ***CPP/FCSm*** (more details in Section 2.2)- still exhibit a cumulative advantage scaling behaviour. We will show in this paper that this is not the case.

Another important indicator is the mean impact level of the journals used by a research group for their publications. The second part of this paper will particularly focus on the problem of possible journal-dependent cumulative advantage. In the context of research performance studies, the discussion on the use of the journal impact factor for evaluation studies regularly flares up. A recent example is the discussion in *Nature* (initiated by the paper of Lawrence 2003) in which researchers, referring to the work of Seglen (1992, 1994), fulminate against the supposed dominant role of journal status and journal impact factors in the present-day life of a scientist. In his study, Seglen concluded that the citedness of publications is not affected by journal status, and that, therefore, ‘certain journals have a high impact simply because they publish high-impact articles’. Thus, the use of journal impact as an indicator for research performance evaluation is inappropriate as the skewed distributions result in poor correlations between article citedness and journal impact. In his work, journal impact was restricted to the ISI journal impact factor only, and did not consider the more sophisticated types of journal impact indicators used in this study (Moed and Van Leeuwen 1995, 1996).

An important finding of Seglen was the poor correlation between the impact of publications and journal impact, both for the whole publication set as well as for individual authors, at *the level of individual publications*. However, grouping publications in classes of journal impact yielded a high correlation between publication and journal impact. But this higher aggregation is determined by journal impact classes, and not by a ‘natural’ higher aggregation level such as a research group. We will show in this paper a quite significant correlation between the average number of citations per publication (publication citedness) of research groups, and the average journal impact of these groups.

A first step in a more differentiated approach to the analysis of statistical properties of bibliometric characteristics of research groups is the distinction between levels of impact of research. Within a set of all 157 university chemistry groups in the Netherlands and a set of 65 medical research groups within a university, we distinguish in our analysis between *top-performance* and *lower performance* groups. The structure of this paper is as follows. In Section 2 we discuss the data material of both sets of research groups, the application of the method, and calculation of indicators. Section 3 addresses the statistical analysis and discusses the results of the analysis. Finally, in Section 4 we summarize the main outcomes of this study.

## 2. Data Material and Bibliometric Indicators

### 2.1 The two datasets

We studied the statistics of bibliometric indicators on the basis of two large sets of publications (as far as published in journals covered by the Citation Index, 'CI publications'[1]), one covering all academic chemistry research in a country (Netherlands) for a 10-years period (1991-2000), and the second covering all research groups in a large medical institution (Leiden) for a period of 12 years. This material is quite unique, as to our knowledge no such compilations of very accurately verified publication sets on a large scale are used for statistical analysis of the characteristics of the indicators at the research group level.

We stress again that the research level is the most important 'work floor entity' in science. However, data at the research group level are by far a trivial matter because 'externally stored' information (such as CI-data on author names, addresses, journals, fields, citations, etc.) has to be combined carefully with 'internally stored' data, i.e., data only available from the institutions that are the 'target' of the analysis. In other words, there are no data on actual research groups available externally like the availability of data on the level of the individual scientist. The only possibility to study the bibliometric characteristics of research groups with 'external data' would be to use the address information within the main organization, for instance 'Department of Biochemistry' of a specific university. However, the delineation of departments or university groups through externally available data such as the address information in the CI databases, is very problematic. We refer for a thorough discussion of this problem to Van Raan (2005b). As indicated above, the data used in this study are the results of evaluation studies and are therefore based on strict verification procedures in close collaboration with the evaluated groups.

The first set concerns all journal articles of all university research groups in chemistry and chemical engineering in the Netherlands (NL). Thus, publications such as reports and books or book chapters are not taken into account. However, for chemistry research groups the focus on papers published in CI-covered journals generally provides a very good representation of the scientific output (VSNU 2002). These ('CI-') publications were collected as part of a large evaluation study conducted by the Association of Universities in the Netherlands. For a detailed discussion of the evaluation procedure and the results we refer to the evaluation report (VSNU 2002). In the framework of this evaluation study, we performed an extensive bibliometric analysis as a support to the international peer committee (van Leeuwen *et al* 2002). The time period covered is 1991-2000 for both publications and citations received by these publications. In total, the analysis involves 700 senior researchers and covers about 18,000 publications and 175,000 citations (excluding self-citations) of 157 chemistry groups.

The indicators are calculated on the basis of the 'total block analysis', which means that publications are counted for the entire 10-year period from 1991-2000 and citations are counted up to and including 2000 (e.g., for publications from 1991, citations are counted in the period 1991-2000, and for publications from 2000, citations are counted only in 2000).

[1] Thomson Scientific, the former Institute for Scientific Information (ISI) in Philadelphia, is the producer and publisher of the Science Citation Index, the Social Science Citation Index, the Arts & Humanities Citation Index, and the 'specialty' citation indexes (CompuMath, Biochemistry and Biophysics, Biotechnology, Chemistry, Material Science, Neurosciences). Throughout this paper we use the term 'CI' (Citation Index) for the above set of databases.

The universities covered by this evaluation study are Leiden, Utrecht, Groningen, Amsterdam UvA, Amsterdam VU, Nijmegen, Delft, Eindhoven, Enschede (Twente), and Wageningen. All fields within the chemistry were covered by this set of university groups, the main fields being analytical chemistry, spectroscopy and microscopy; computational and theoretical chemistry, physical chemistry; catalysis; inorganic chemistry; organic and bio-organic chemistry; biochemistry, microbiology and biochemical engineering; polymer science and technology; materials science; chemical engineering.

The second set concerns all publications (again as far as published in journals covered by the Citation Index, 'CI publications') of all research groups in the Leiden University Medical Center (LUMC). Also in this case, the focus on papers published in CI-covered journals generally provides a very good representation of the scientific output. These publications were collected as part of an internal Leiden evaluation study. In the framework of this evaluation study, we performed a detailed bibliometric analysis as a support to the LUMC research evaluation board. Details of the results are available from the author of this paper. The time period covered is 1990-2001 for both publications and the citation received by these publications. The citation counting procedure is the same as for the chemistry groups. In total, the analysis involves 400 senior researchers and covers about 10,000 publications and 185,000 citations (excluding self-citations) of 65 medical groups. The LUMC is a large clinical and basic research organization of high international reputation. Practically all fields of medical research are present, ranging from molecular cell biology to oncological surgery, and from organ transplantation to T-cell immune response research.

## 2.2. The bibliometric analysis and applied indicators

We apply the CWTS standard bibliometric indicators. Here only 'external' citations, i.e., citations corrected for self-citations, are taken into account. An overview of these indicators in given in the textbox in this section. In particular, we draw the attention to the definition of our journal impact indicator, ***JCSm***. For a detailed discussion we refer to Van Raan (1996, 2004, 2005d).

***CWTS Standard Bibliometric Indicators*:**

- Number of publications (***P***) in CI-covered journals of the research group in the entire period;
- Number of citations received by ***P*** during the entire period, with and without self-citations (***Ci*** and ***C***);
- Average number of citations per publication, with and without self-citations (***CPPi*** and ***CPP***);
- Percentage of publications not cited (in the given time period), ***Pnc***;
- Journal-based worldwide average impact as an international reference level for the research group (***JCS***, journal citation score), without self-citations (on this world-wide scale!), in the case of more than one journal (as almost always) we use the weighted average ***JCSm***; for the calculation of ***JCSm*** the same publication and citation counting procedure, time windows, and article types are used as in the case of ***CPP***;
- Field-based worldwide average impact as an international reference level for the research group (***FCS***, field citation score), without self-citations (on this world-wide scale!) in the case of more than one field (as almost always) we use the weighted average ***FCSm***; for the calculation of ***FCSm*** the same publication and citation counting procedure, time windows, and article types are used as in the case of ***CPP***;
- Comparison of the actually received international impact of the research group with the world-wide average based on ***JCSm*** as a standard, without self-citations, indicator ***CPP/JCSm***;
- Comparison of the actually received international impact of the research group with the world-wide average based on ***FCSm*** as a standard, without self-citations, indicator ***CPP/FCSm***;
- Ratio ***JCSm/FCSm*** as journal-level indicator, e.g., to answer the question: is the research group publishing in top or in sub-top (in terms of 'citedness') journals?
- Percentage of self-citations of the research group, ***SelfCit.***

In Table 1 we show as an example the results of our bibliometric analysis for the most important indicators for all 12 chemistry research groups of one of the ten universities ('Univ A'). Also the quality judgement ***Q*** of the international peer committee is indicated. The peers used a three-point scale to judge the research quality of a group: Grade 5 is 'excellent', Grade 4 is 'good', and Grade 3 is 'satisfactory' (VSNU 2002).

***Table 1****: Example of the results of the bibliometric analysis for the chemistry groups*

| ***Research group*** | ***P*** | ***C*** | ***CPP*** | ***JCSm*** | ***FCSm*** | ***CPP/JCSm*** | ***CPP/FCSm*** | ***JCSm/FCSm*** | ***Quality*** |
|---|---|---|---|---|---|---|---|---|---|
| | | | | | | | | | |
| Univ A, 01 | 92 | 554 | 6.02 | 5.76 | 4.33 | 1.05 | 1.39 | 1.33 | 5 |
| Univ A, 02 | 69 | 536 | 7.77 | 5.12 | 2.98 | 1.52 | 2.61 | 1.72 | 4 |
| Univ A, 03 | 129 | 3,780 | 29.30 | 17.20 | 11.86 | 1.70 | 2.47 | 1.45 | 5 |
| Univ A, 04 | 80 | 725 | 9.06 | 8.06 | 6.25 | 1.12 | 1.45 | 1.29 | 4 |
| Univ A, 05 | 188 | 1,488 | 7.91 | 8.76 | 5.31 | 0.90 | 1.49 | 1.65 | 5 |
| Univ A, 06 | 52 | 424 | 8.15 | 6.27 | 3.56 | 1.30 | 2.29 | 1.76 | 4 |
| Univ A, 07 | 52 | 362 | 6.96 | 4.51 | 5.01 | 1.54 | 1.39 | 0.90 | 3 |
| Univ A, 08 | 171 | 1,646 | 9.63 | 6.45 | 4.36 | 1.49 | 2.21 | 1.48 | 5 |
| Univ A, 09 | 132 | 2,581 | 19.55 | 15.22 | 11.71 | 1.28 | 1.67 | 1.30 | 4 |
| Univ A, 10 | 119 | 2,815 | 23.66 | 22.23 | 14.25 | 1.06 | 1.66 | 1.56 | 4 |
| Univ A, 11 | 141 | 1,630 | 11.56 | 17.83 | 12.30 | 0.65 | 0.94 | 1.45 | 4 |
| Univ A, 12 | 102 | 1,025 | 10.05 | 10.48 | 7.18 | 0.96 | 1.40 | 1.46 | 5 |

We applied the same bibliometric indicators to the medical research groups. An example of the results (first 10 groups) is presented in Table 2. Thus, the results of both cases are based on a strictly consistent methodology and thus they are directly comparable. Only in the LUMC (medical) case we used a somewhat longer period (12 years) as compared to the chemistry case (10 years). The LUMC research evaluation board did not call in an international peer committee, so no quality judgement data as in the case of the chemical research groups are available. We added two further standard indicators, the percentage of not-cited publications (***Pnc***) and the percentage of self-citations (***Scit***).

***Table 2****: Example of the results of the bibliometric analysis for the medical groups*

| ***Research group*** | ***P*** | ***C*** | ***CPP*** | ***JCSm*** | ***FCSm*** | ***CPP/JCSm*** | ***CPP/FCSm*** | ***JCSm/FCSm*** | ***Pnc*** | ***Scit*** |
|---|---|---|---|---|---|---|---|---|---|---|
| | | | | | | | | | | |
| LU 01 | 117 | 1,836 | 15.69 | 12.20 | 11.08 | 1.29 | 1.42 | 1.10 | 11% | 20% |
| LU 02 | 197 | 3,587 | 18.21 | 14.28 | 14.75 | 1.28 | 1.23 | 0.97 | 11% | 21% |
| LU 03 | 46 | 449 | 9.76 | 14.55 | 8.78 | 0.67 | 1.11 | 1.66 | 20% | 23% |
| LU 04 | 560 | 16,906 | 30.19 | 25.22 | 15.29 | 1.20 | 1.97 | 1.65 | 10% | 19% |
| LU 05 | 423 | 17,144 | 40.53 | 29.60 | 16.85 | 1.37 | 2.41 | 1.76 | 6% | 21% |
| LU 06 | 369 | 13,454 | 36.46 | 30.34 | 17.54 | 1.20 | 2.08 | 1.73 | 6% | 19% |
| LU 07 | 91 | 1,036 | 11.38 | 11.91 | 7.72 | 0.96 | 1.47 | 1.54 | 15% | 22% |
| LU 08 | 95 | 554 | 5.83 | 6.52 | 5.80 | 0.89 | 1.01 | 1.13 | 22% | 33% |
| LU 09 | 52 | 334 | 6.42 | 6.98 | 8.00 | 0.92 | 0.80 | 0.87 | 23% | 33% |
| LU 10 | 512 | 5,729 | 11.19 | 8.70 | 6.44 | 1.29 | 1.74 | 1.35 | 22% | 17% |

Tables 1 and 2 make clear that our indicator calculations allow a statistical analysis of these indicators for both sets of research groups. Of the above indicators, we regard the internationally standardized (field-normalized) impact indicator ***CPP/FCSm*** as our 'crown' indicator. This indicator enables us to observe immediately whether the performance of a research group is significantly far below (indicator value < 0.5), below (indicator value 0.5 - 0.8), around (0.8 - 1.2), above (1.2 – 1.5), or far above (>1.5) the international (western world dominated) impact standard of the field. Particularly with a ***CPP/FCSm*** value above 1.5, groups can be considered as scientifically strong, a value above 2 indicates a very strong group, and above 3 the groups can be, generally, considered as really excellent and comparable to top-groups at the best US universities (van Raan 1996, 2000, 2004).

The ***CPP/FCSm*** indicator generally correlates well with the quality judgement of the peers. Studies of larger-scale evaluation procedures in which empirical material is available with data on both peer judgment as well as bibliometric indicators are quite rare. We refer to Rinia *et al* (1998) for a comparison of bibliometric assessment based on various indicators with peer review judgment in condensed matter physics, and to Rinia *et al* (2001) for a study of the influence of interdisciplinarity on peer review in comparison with bibliometric indicators. Our analysis deals with primarily the indicators ***P***, ***C***, ***CPP***, ***JCSm***, ***CPP/FCSm***, and ***Pnc***.

The set of chemistry groups and the set of medical groups differ in, particularly for this study, relevant aspects. The chemistry groups are from ten different universities, they have grown more or less 'naturally', and they are not subject to one specific research policy strategy as all these ten universities have their own priorities. The medical groups, however, are all within one large institution. Although they also can be considered as having a 'natural' basis as a research group around one or two full professors, these groups are at the same time influenced by the policy of the LUMC as a whole. For instance, close mutual collaboration and the availability of the best people and facilities of a wide range of groups in the same location may enhance performance.

## 3. Results and Discussion

### 3.1 Scale-dependent cumulative advantage of impact

In our earlier study (van Raan 2005a) on the statistical behaviour of bibliometric indicators we showed how a specific collection of publications (namely, a research group) is characterized in terms of the correlation between 'size' (the total number of publications ***P*** of a specific research group[2]) and the total number of citations received by this group in a given period of time, ***C***. This relation for all 157 chemistry research groups is presented in Fig. 1a, and for the 65 medical research groups in Fig. 1b. These figures show us that this relation *on the aggregation level of research groups* is described with reasonable significance (coefficient of determination of the fitted regression is $R^2 = 0.69$ and 0.86, respectively) by a power law:

$\boldsymbol{C}(\boldsymbol{P}) = 2.04\, \boldsymbol{P}^{1.25}$, for the chemistry research groups

$\boldsymbol{C}(\boldsymbol{P}) = 0.56\, \boldsymbol{P}^{1.60}$, for the medical research groups

[2] The number of publications is a valid measure of size in the statistical context described in this paper. It is, however, a proxy for the 'real size' of a research group in terms number of, for instance, *staff full time equivalents* (fte) available for research.

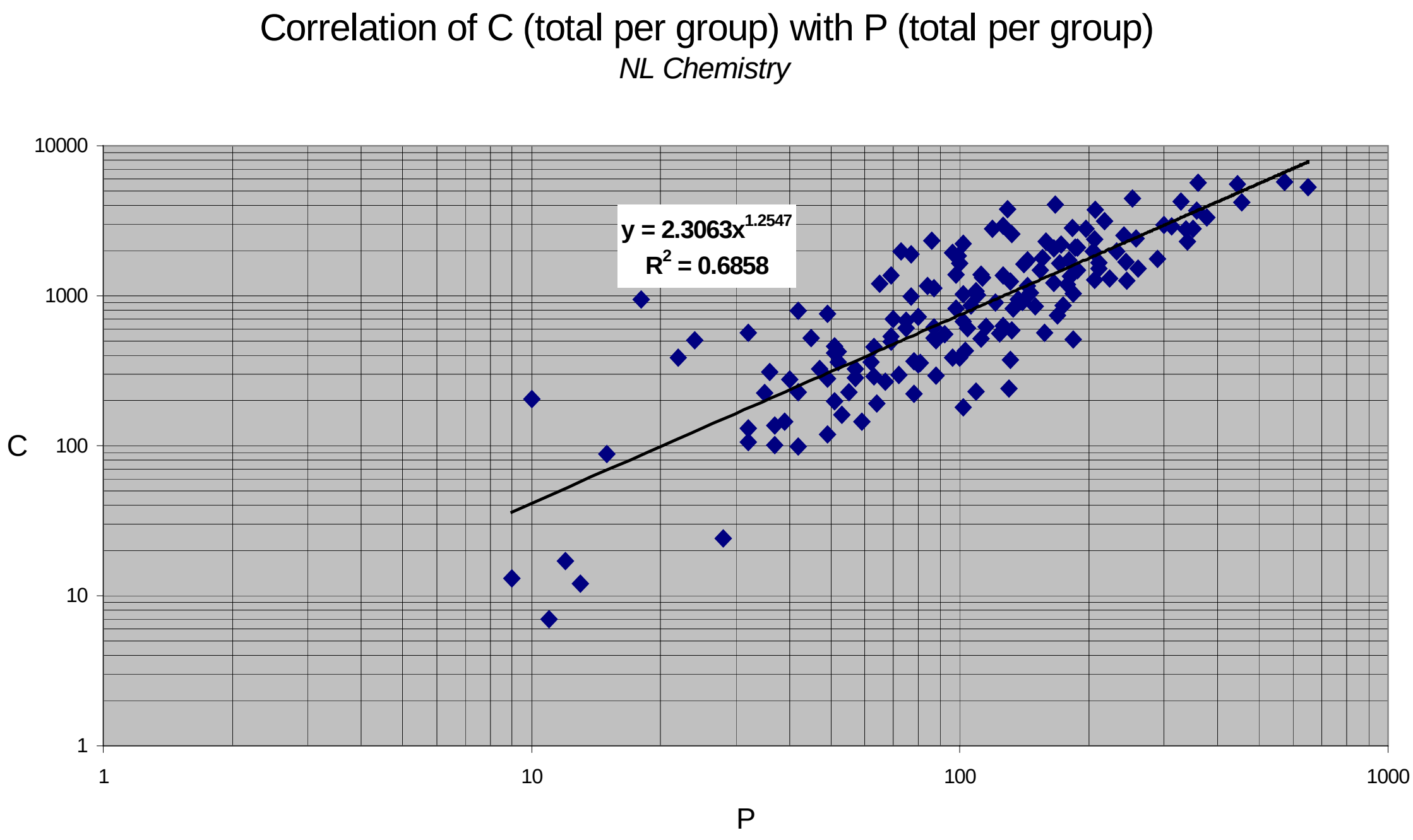

***Fig. 1a:*** *Correlation of the number of citations (**C**) received per research group with the number of publications (**P**) of these groups, for all chemistry groups*.

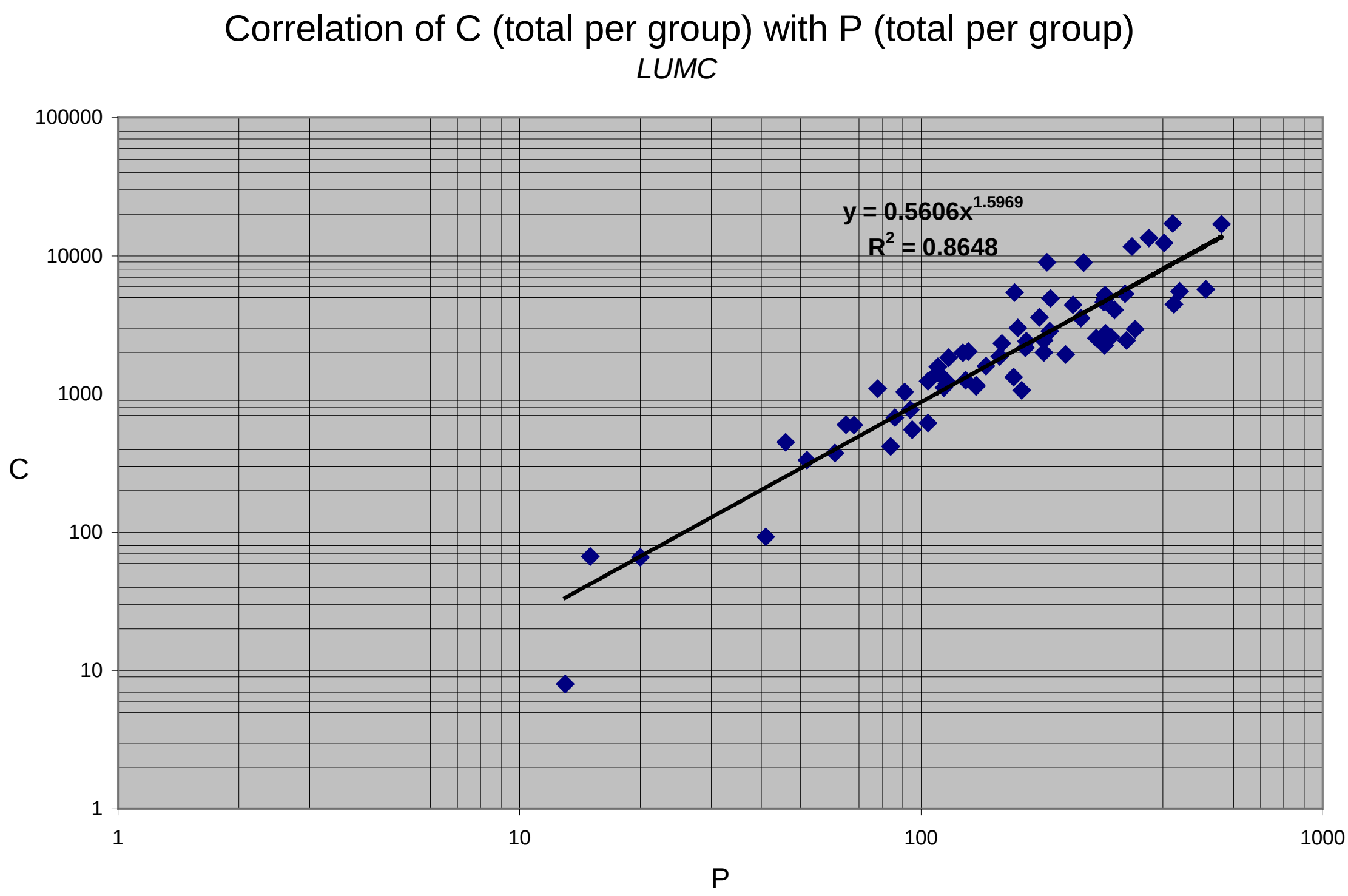

***Fig. 1b:*** *Correlation of the number of citations (**C**) received per research group with the number of publications (**P**) of these groups, for all medical research (LUMC) groups.*

We observe that the size of groups leads to a 'cumulative advantage' (with exponents +1.25 and +1.60, respectively) for the number of citations received by these groups. Thus the earlier discussed assumption of Katz (1999, 2000) that the 'Matthew effect' also works in a sufficiently large set of research groups, is confirmed. But a remarkable further finding is that this scale-dependent cumulative advantage is considerably larger for the LUMC groups. As discussed in Section 2.2, the medical groups are different from the chemistry groups in organisational aspects that may enhance performance. However, further study is necessary to find out whether these differences are also or even to a major extent due to disciplinary characteristics.

In our earlier study, we stressed that in the context of research performance analysis studies, scale-dependent 'corrections' (on the basis of number of publications) of measured impact (on the basis of citations) will lead to an unreasonable levelling off of the impact indicators at the level of research groups. This is because size can be regarded in many cases as an intrinsic characteristic of performance: top-research groups attract promising people, and thus these groups will have an advantage to grow. Therefore, it is of crucial importance to investigate the above correlation for those subsets of the entire set that represent clearly *differences in research performance*. For the chemistry groups we have, next to the bibliometric indicators, also the peer review results, as discussed in Section 2.2.

We first created within the entire set of chemistry groups two subsets on the basis op the quality judgement by peers. One subset with 39 'top-performance' groups, these groups received the highest judgment 'excellent' (***Q*** = 5), and another subset with 30 lower performance groups, these groups that received the lowest judgment 'satisfactory' (***Q*** = 3). The results are given in Fig. 2.

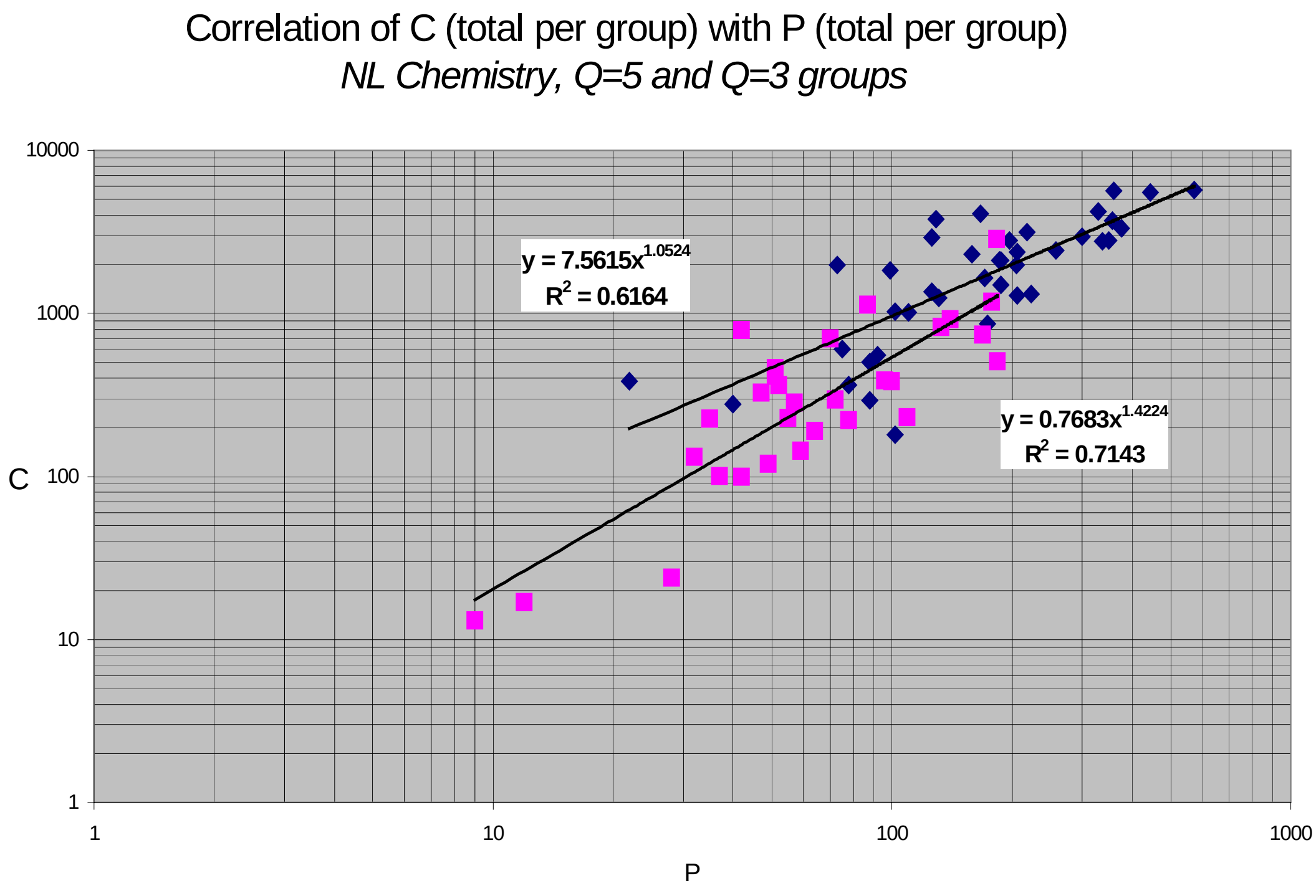

***Fig. 2:*** *Correlation of the number of citations (**C**) received per chemistry research group with the number of publications (**P**), for the top-performance groups (Q=5, indicated with diamonds), and for the lower performance groups (Q=3, indicated with squares).*

We observe a striking difference between the two subsets. The top-performance groups generally have more total citations for a given size in terms of ***P***, but the 'cumulative advantage' is considerably less (in fact almost not existing, exponent 1.05) than for the lower performance groups (exponent 1.42). A next step is to create subsets within the entire set of chemistry research groups and medical research groups on the basis of our research performance (field-normalized impact) indicator ***CPP/FCSm***.

We created the following subsets: groups belonging to the top-10%, top-20%, and top-50%, as well as to the bottom-10%, bottom-20%, and bottom-50% of the ***CPP/FCSm*** distribution. We present the results of the correlation measurement in Figs. 3a and 3b (top-10% compared to bottom-10%, chemistry and medical groups, respectively), 3c and 3d (top-20% compared to bottom-20%, chemistry and medical groups, respectively), and 3e and 3f (top-50% compared to bottom-50%, chemistry and medical groups, respectively).

We again clearly observe the differences between the two 'opposite' subsets. And similar to the observations in Fig. 2 based on peer-review quality ratings, we notice that the top-performance groups generally have more total citations for a given size in terms of ***P*** (which is, of course, to be expected), but that the cumulative advantage is considerably less for the top-performers than for the lower performance groups. As we have differentiated between top and lower performance in a gradual way (top/bottom 10%, 20%, and 50%), we are able to study the correlation of ***C*** with ***P*** and possible scale effects (size-dependent cumulative advantage) in more detail. The results (based on the observations in Figs. 3a-f) are presented in Table 3.

***Table 3****: Power law exponent **α** of the correlation of **C** with **P** for the two sets of groups, in the indicated modalities. The differences in **α** between the set of chemistry research groups and the set of medical research groups is given by **Δα**(M,C); the difference between the top and bottom modalities (see text) by **Δα**(b,t).*

| | **Chemistry Groups** | **Medical Groups** | ***Δα****(M,C)* |
|---|---|---|---|
| **top 10%** | 0.72 | 1.13 | *0.41* |
| **bottom 10%** | 1.44 | 1.75 | *0.31* |
| ***Δα****(b,t)* | *0.72* | *0.62* | |
| | | | |
| **top 20%** | 0.90 | 1.39 | *0.49* |
| **bottom 20%** | 1.46 | 1.64 | *0.18* |
| ***Δα****(b,t)* | *0.56* | *0.25* | |
| | | | |
| **top 50%** | 1.06 | 1.54 | *0.48* |
| **bottom 50%** | 1.44 | 1.55 | *0.11* |
| ***Δα****(b,t)* | *0.38* | *0.01* | |

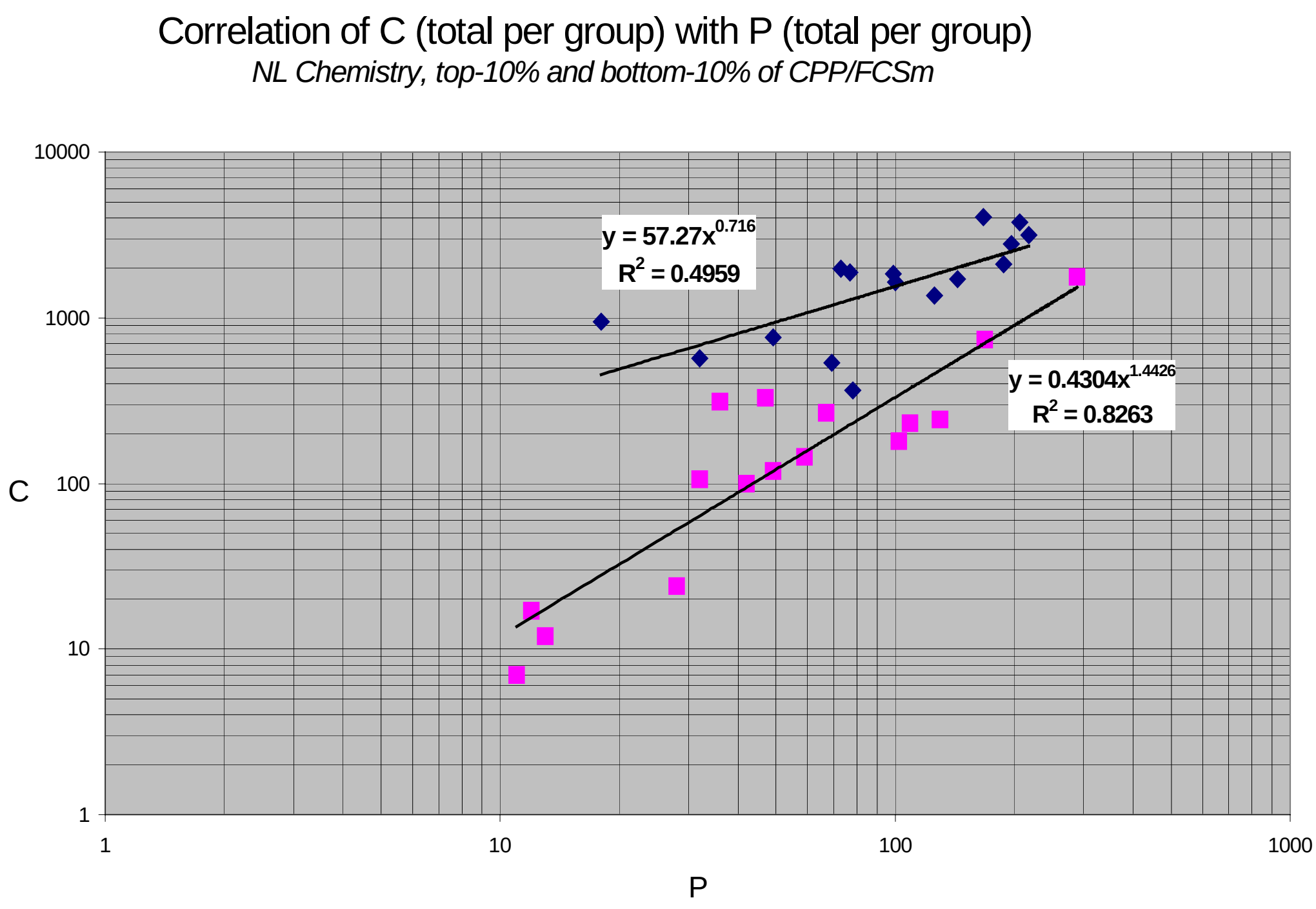

***Fig. 3a:*** *Correlation of the number of citations (**C**) received per chemistry research group with the number of publications (**P**), for the top-10% (of **CPP/FCSm**) groups (indicated with diamonds), and for the bottom-10% groups (indicated with squares).*

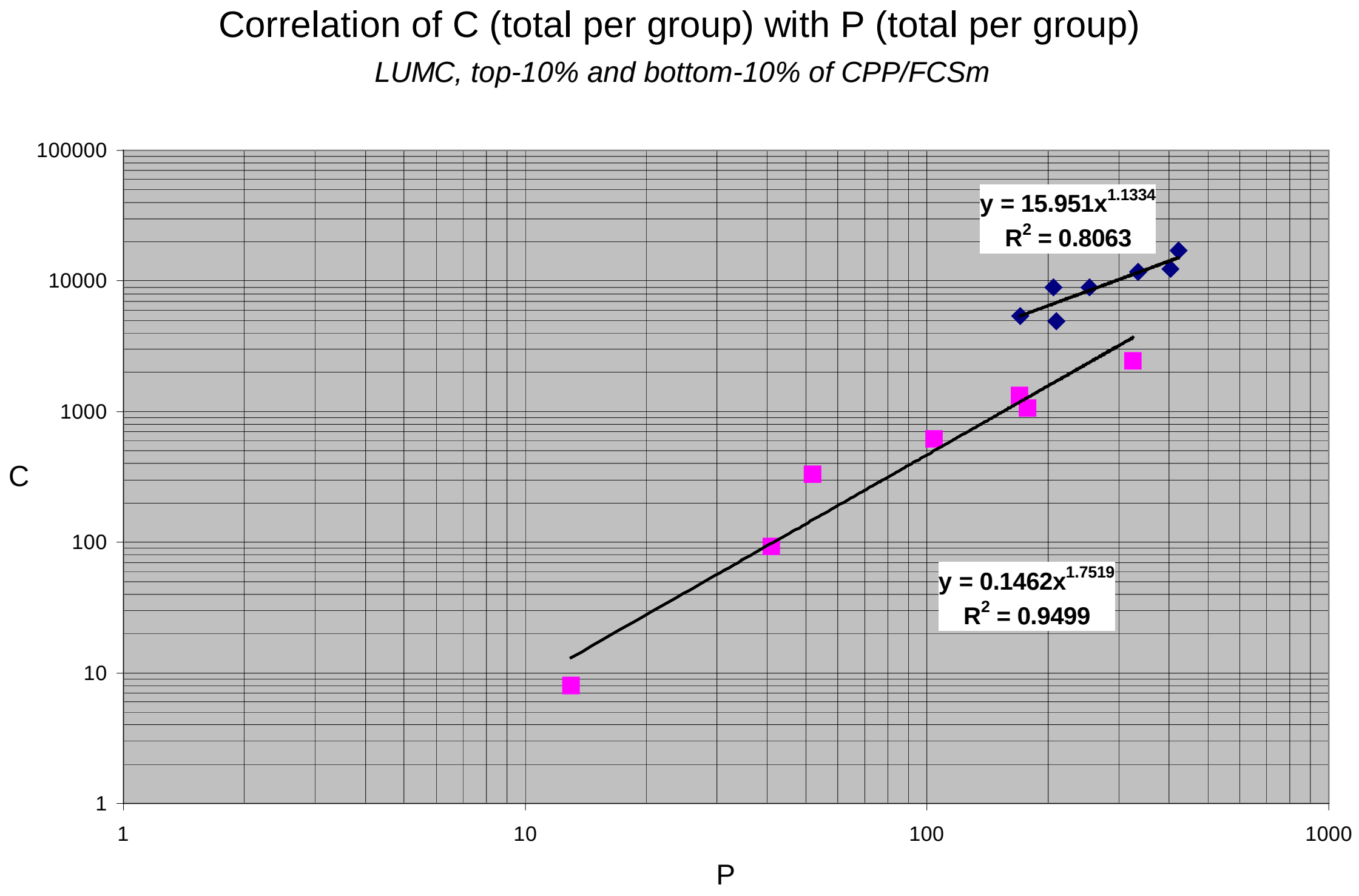

***Fig. 3b:*** *Correlation of the number of citations (**C**) received per medical (LUMC) research group with the number of publications (**P**), for the top-10% (of **CPP/FCSm**) groups (indicated with diamonds), and for the bottom-10% groups (indicated with squares).*

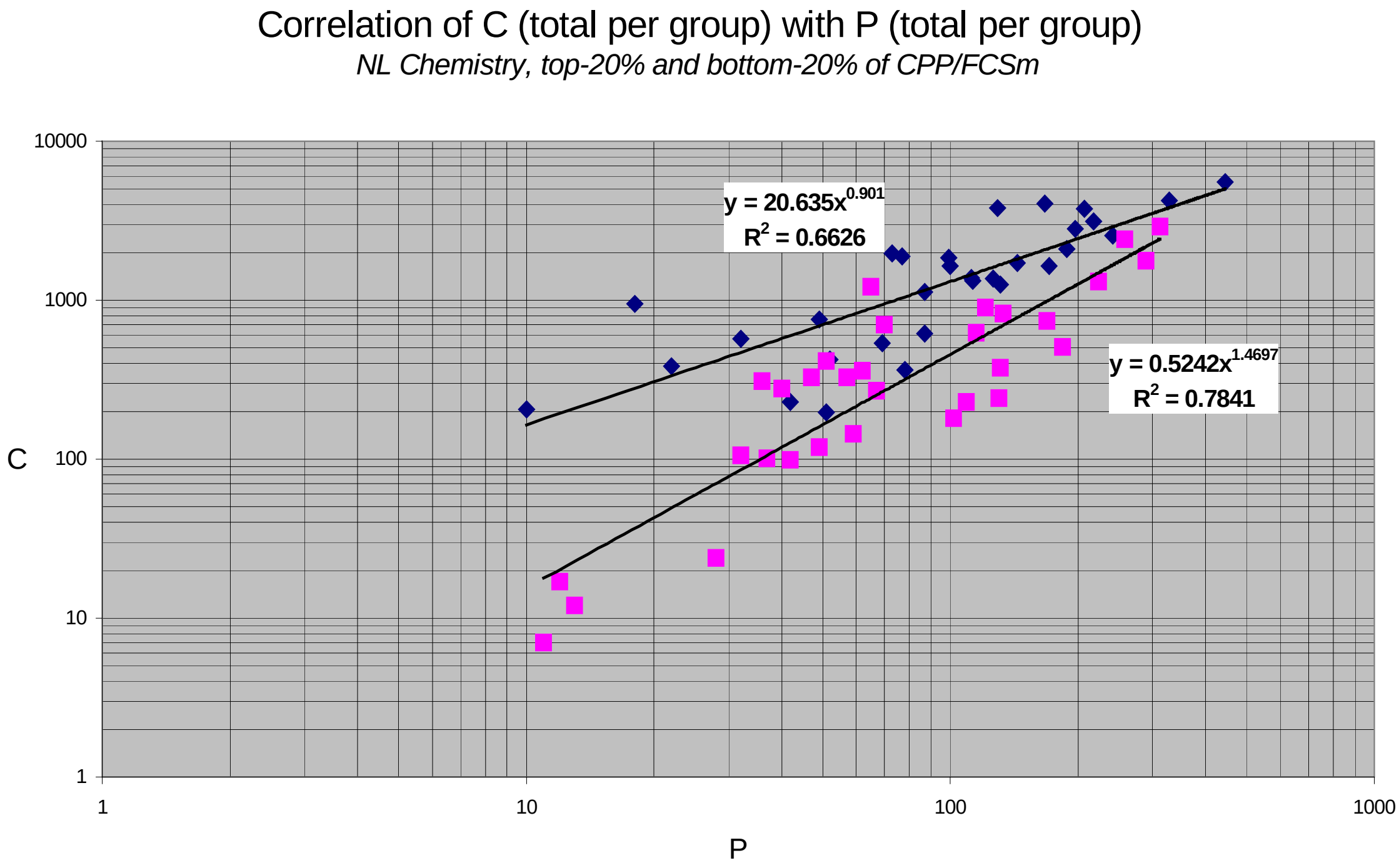

***Fig. 3c:*** *Correlation of the number of citations (**C**) received per chemistry research group with the number of publications (**P**), for the top-20% (of **CPP/FCSm**) groups (indicated with diamonds), and for the bottom-20% groups (indicated with squares).*

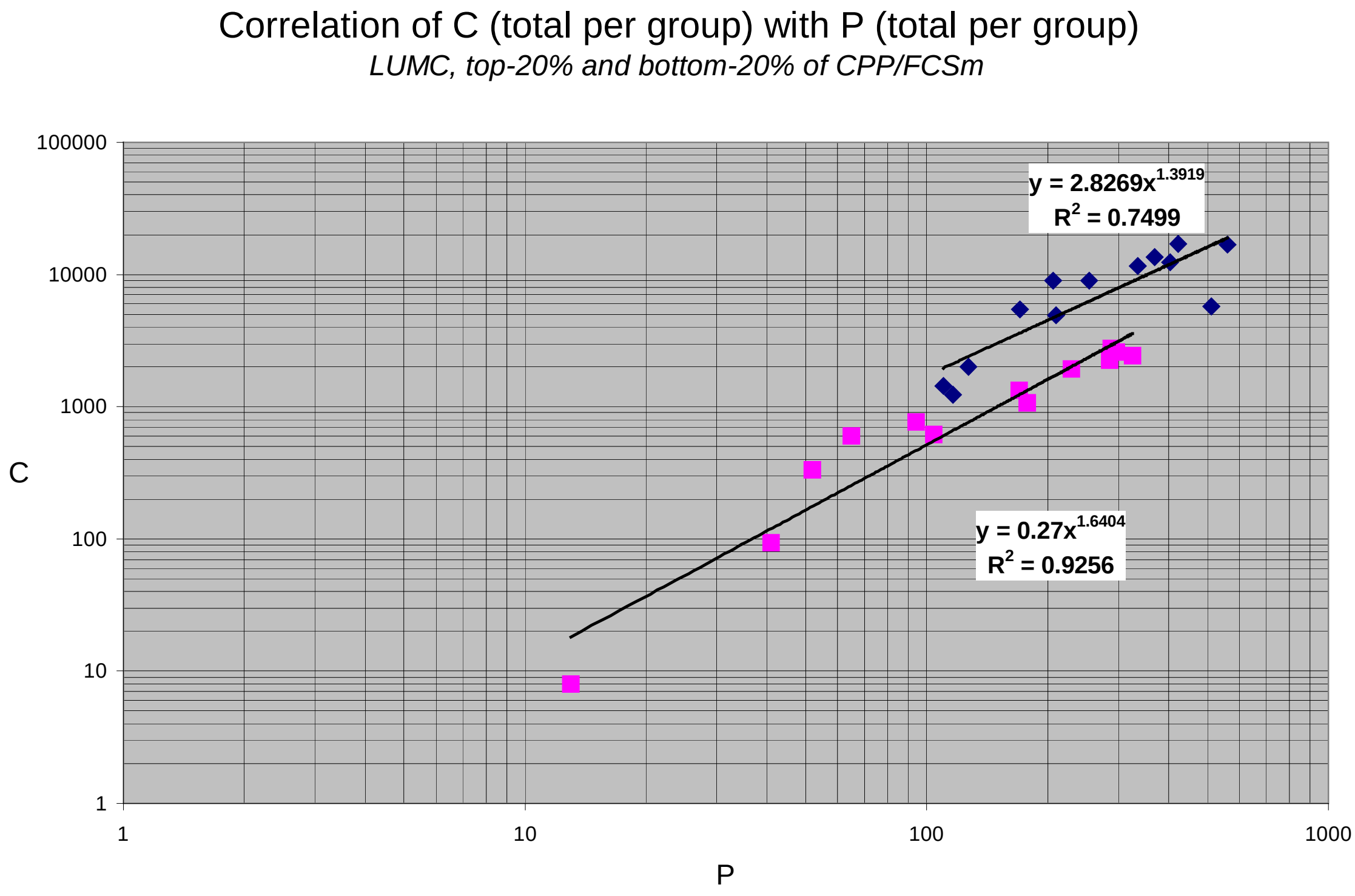

***Fig. 3d:*** *Correlation of the number of citations (**C**) received per medical (LUMC) research group with the number of publications (**P**), for the top-20% (of **CPP/FCSm**) groups (indicated with diamonds), and for the bottom-20 groups (indicated with squares).*

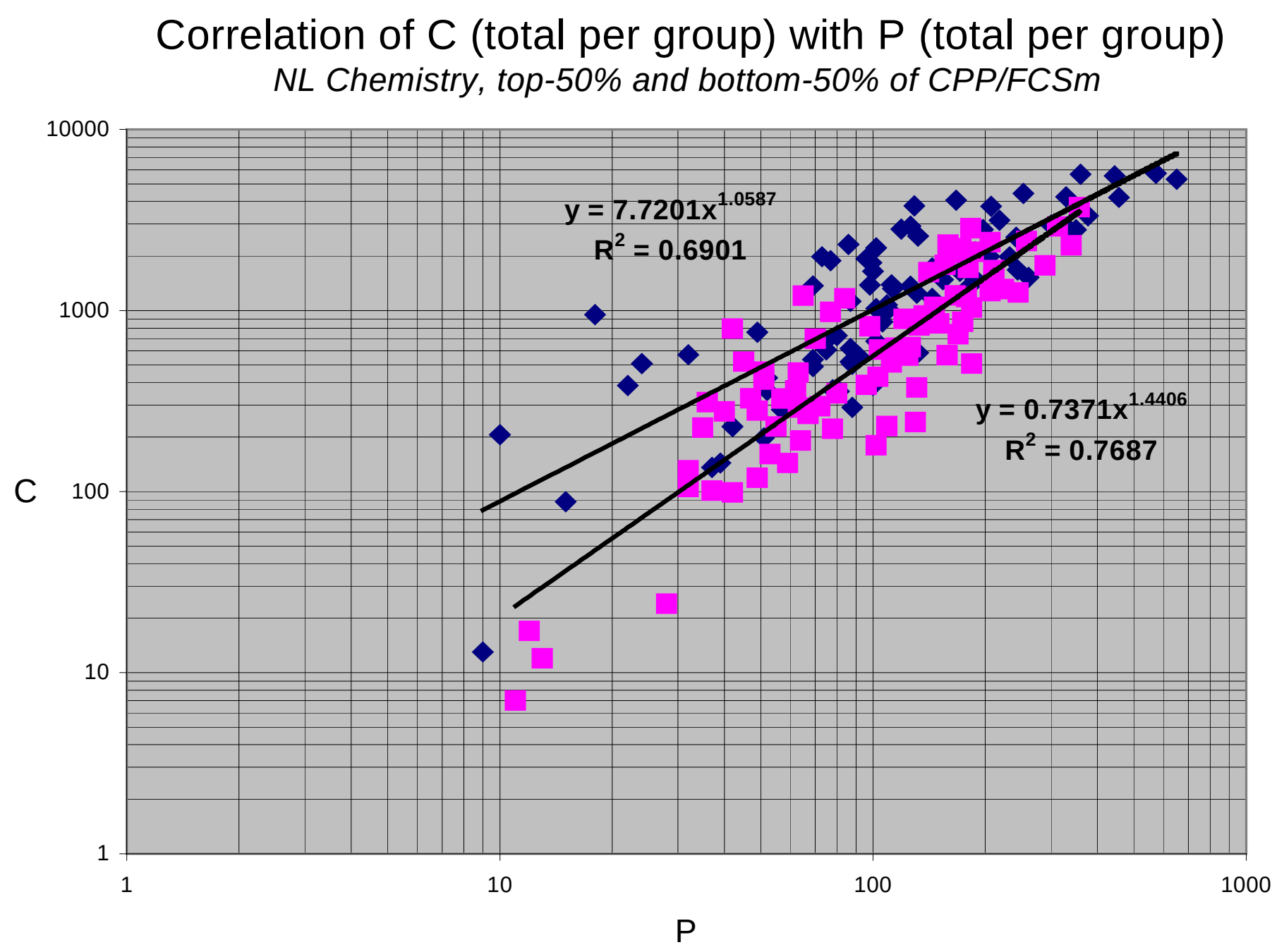

***Fig. 3e:*** *Correlation of the number of citations (**C**) received per chemistry research group with the number of publications (**P**), for the top-50% (of **CPP/FCSm**) groups (indicated with diamonds), and for the bottom-50% groups (indicated with squares).*

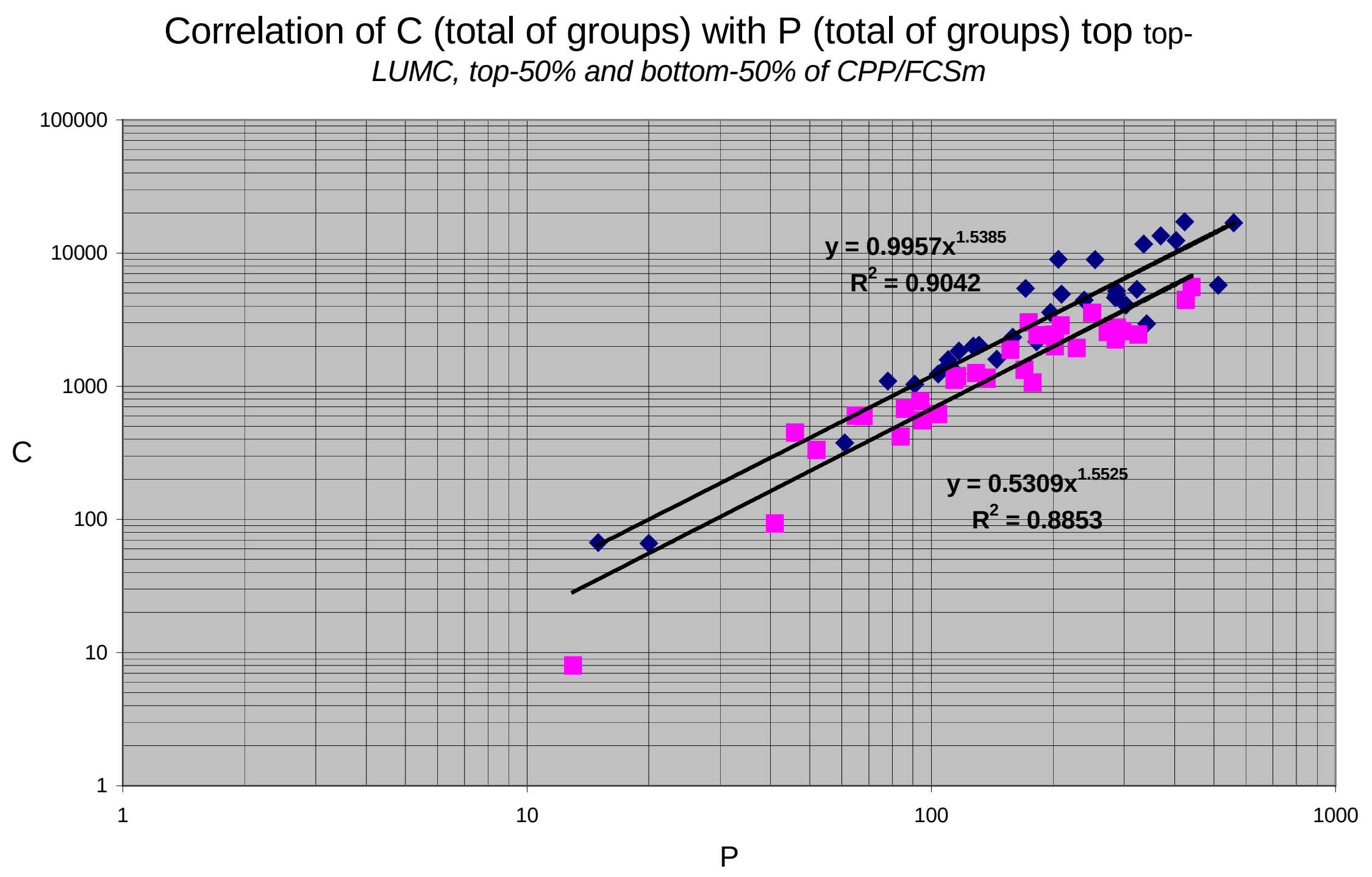

***Fig. 3f:*** *Correlation of the number of citations (**C**) received per medical (LUMC) research group with the number of publications (**P**), for the top-50% (of **CPP/FCSm**) groups (indicated with diamonds), and for the bottom-50% groups (indicated with squares).*

Table 3 shows us the following findings. First, the medical research groups have a stronger 'advantage' with size (***P***) than the chemistry research groups, particularly for top-groups. For the 'bottom' the difference in advantage $\Delta\alpha$(M,C) between medical and chemistry groups is smaller.

Second, for the medical research groups the difference in advantage between top- and bottom-groups $\Delta\alpha$(b,t) is smaller than for the chemistry groups. Third, the top-10% and top-20% of the chemistry groups do not have a cumulative advantage (i.e., an exponent > 1), and for the top-50% there is just a very small cumulative advantage. For the medical research groups, the cumulative advantage is clearly visible in all modalities. Fourth, and the most intriguing observation, we observe, as noticed earlier, that for both the chemistry as well as the medical research groups, the 'bottom' groups profit more than the top-groups (with an exception for the bottom-50% of the medical research groups).

This latter phenomenon has the consequence that for a specific size (***P***), top-10% and bottom-10% have almost the same 'strength' (***C***), see for instance Fig 3e, from about ***P*** = 400. Research groups are active within a specific theme or subfield of a discipline; hence the total number of 'available' citations will get 'exhausted'. Thus, a saturation of cumulative advantage is unavoidable simply by finite-size considerations.

### 3.2 Modelling advantages

In the foregoing section we discussed that the most interesting finding so far is that *lower* performance groups have a *larger* scale-dependent cumulative advantage than the top-performance groups. So the crucial question now is: what does size actually do?

In an attempt to understand more precisely what is going on, we have to build a bridge between the 'macro' picture given by the correlation between ***C*** and ***P*** at the level of groups, and the 'micro' picture, particularly the distribution of citations over publications ***p***(***C***) within a group. We use the symbol '***P***' for the total number of publications per group, '***p***' for the distribution functions based on the number of 'individual' publications, and '***pr***' for the distribution function with relative numbers of publications. We stress that self-citations are excluded in this analysis.

Figure 4 shows the ***pr***(***C***) distribution for the largest-20% and the smallest-20% of the entire set of chemistry research groups. We stress that the subsets are not, like in Figs. 3a-f, related to differences in performance, but to size difference. Figure 4 reveals a crucial feature of the distribution. The subset of the smallest-20% has a significantly higher fraction of not-cited publications ***pr***(***0***) as compared to the subset of the largest-20%. The ratio for both subsets is 1.40. So, *larger* groups have *less* not cited publications as compared to smaller groups. As a direct consequence we also see that the largest-20% groups have relatively more publications with 2 to 20 citations, which represents a major part of the entire citation distribution. Thus, we suppose that a possible mechanism for cumulative advantage by group size works through decrease of the not-cited publications in a group and 'promoting' the already cited publications. In other words, size reinforces an 'internal promotion mechanism'. We elaborate this idea further on in this section.

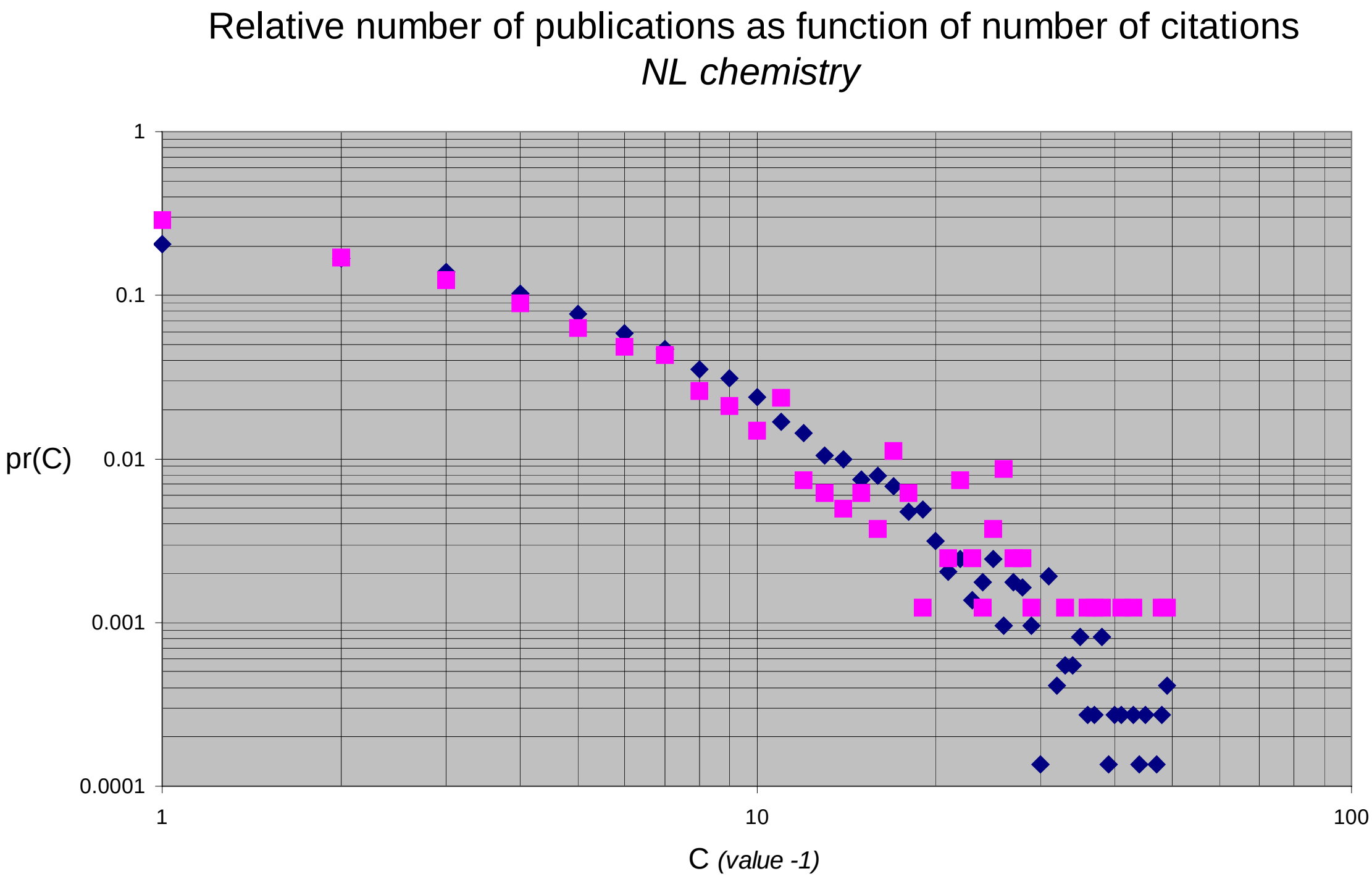

*Fig. 4: Distribution function* ***pr(C)****: relative number of publications as a function of number of citations for two subsets within the total set of 157 chemistry publications: the subset of the groups that belong to the largest 20% (diamonds), and the subset of the groups that belong to the smallest 20% (squares); in order to include* ***C*** *= 0 values in the logarithmic scale, we take on the abscissa value 1 for 0 citations, value 2 for 1 citation, etc.*

As a further additional empirical investigation of this observation we analysed for the medical research groups the correlation of the fraction of not-cited-publications ***Pnc*** of a group (given in Table 2) with the size (in terms of ***P***) of a group. The results are shown in Fig. 5a.

We observe, be it with low significance, that the fraction of not cited publications decreases as function of size in terms of number of publications in a group, which confirms the findings for the chemistry groups on the basis of the distribution functions. But the significance of the correlation is too low for clear results. Thus, as a further step we investigate the correlation of the fraction of not-cited-publications ***Pnc*** of a group with size (in terms of ***P***) with a distinction between top-performance groups and lower-performance groups, similar to our analysis of the ***C***(***P***) correlation. So we split up the entire set of groups presented in Fig 5a, in the top and lower performance groups. The results are presented in Figs. 5b-d.

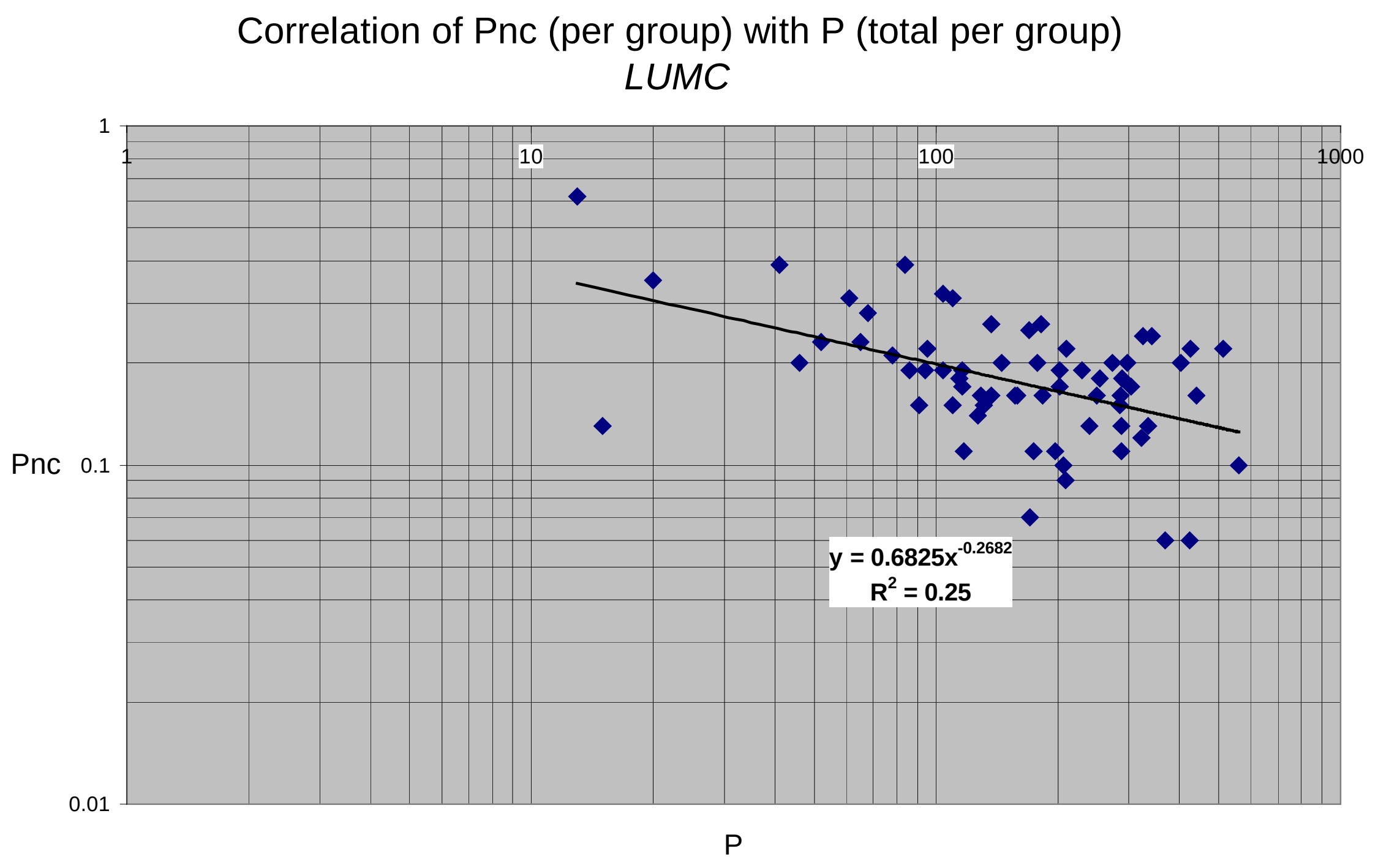

*Fig. 5a: Correlation of the relative number of not cited publications (**Pnc**) of the medical research groups with the number of publications (**P**).*

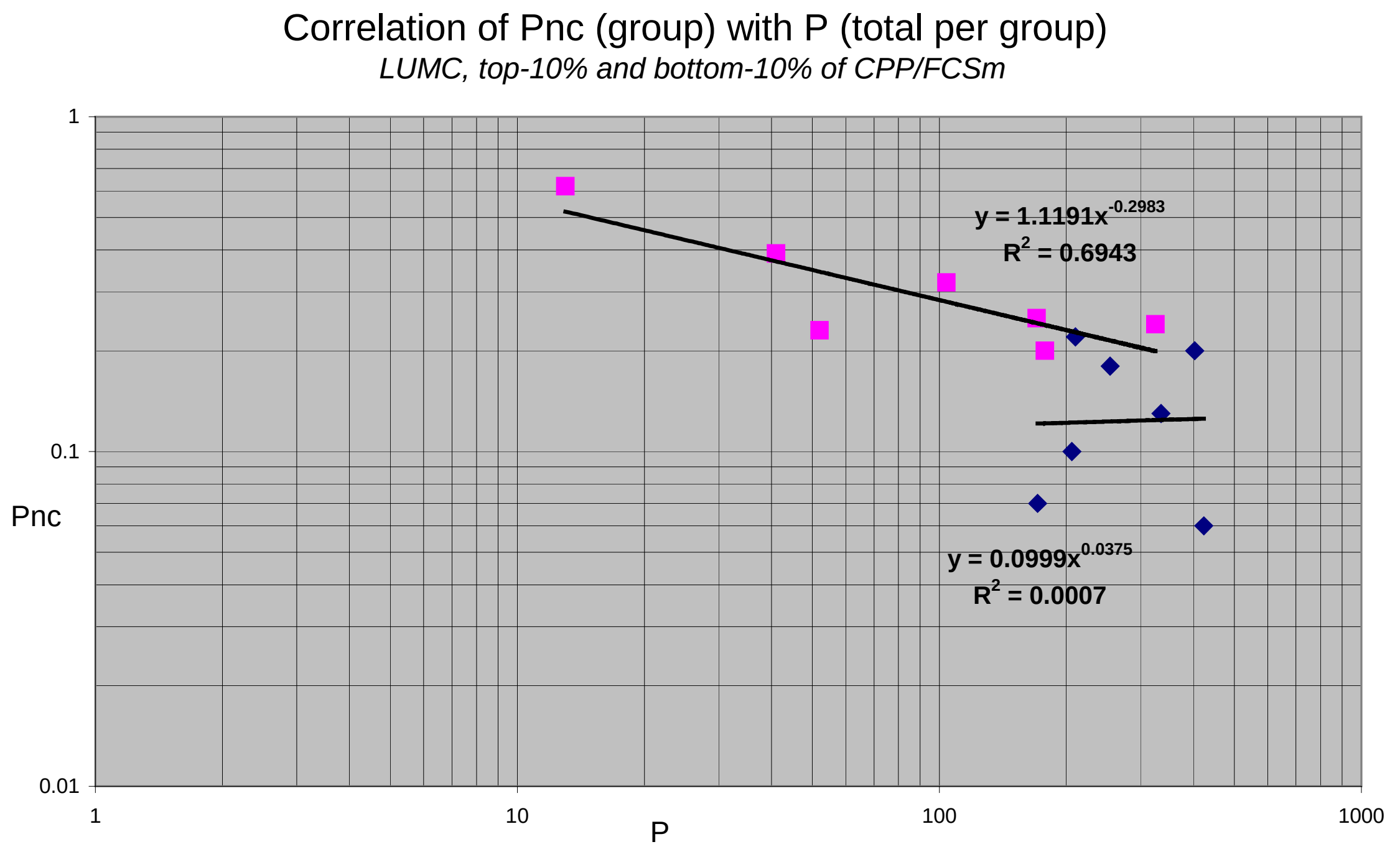

*Fig. 5b: Correlation of the relative number of not cited publications (**Pnc**) of the medical research groups with the number of publications (**P**), for the top-10% (of **CPP/FCSm**) groups (indicated with diamonds), and for the bottom-10% groups (indicated with squares).*

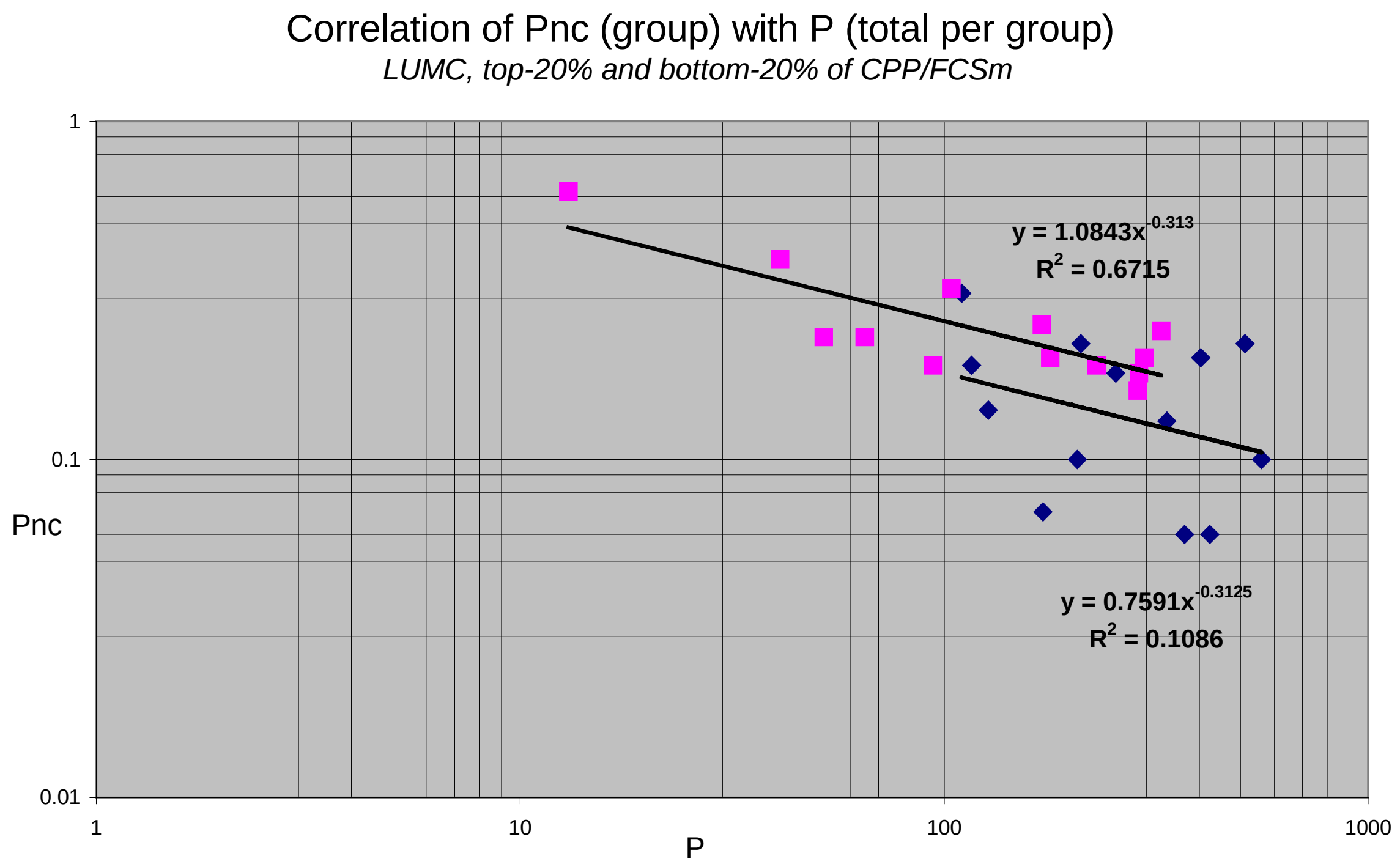

***Fig. 5c:*** *Correlation of the relative number of not cited publications (**Pnc**) of the medical research groups with the number of publications (**P**), for the top-20% (of **CPP/FCSm**) groups (indicated with diamonds), and for the bottom-20% groups (indicated with squares).*

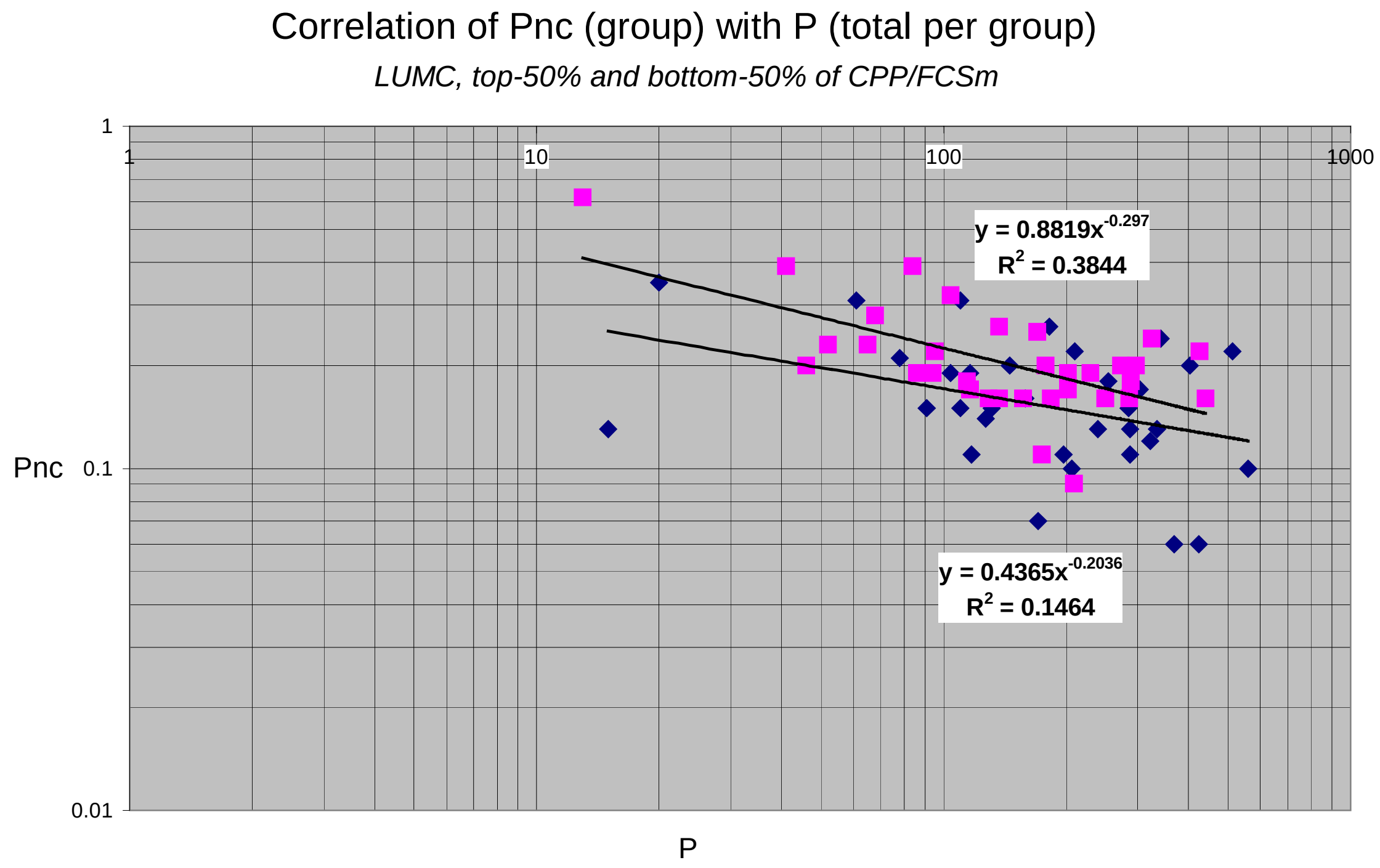

***Fig. 5d:*** *Correlation of the relative number of not cited publications (**Pnc**) of the medical research groups with the number of publications (**P**), for the top-50% (of **CPP/FCSm**) groups (indicated with diamonds), and for the bottom-50% groups (indicated with squares).*

The figures reveal several remarkable features. First, the LUMC top-performance groups are generally the larger ones, i.e., in the right hand side of the correlation function. The correlation with ***Pnc*** is not significant. Second, the lower-performance groups benefit from size, with reasonable significance for the bottom-10% and -20% (Figs. 5b and 5c). In other words, an important observation in this study is that particularly for the lower performance groups, with reasonable significance, the fraction of non-cited publications decreases with size.

Our explanation is that advantage by size works by a mechanism in which the number of not-cited publications is diminished, and that this mechanism is particularly effective for the lower performance groups. We stress again that in our analysis self-citations are excluded!
Thus, the larger the number of publications in a group, the more those publications in the group that otherwise would have remain uncited, are 'promoted'. Most probably this works by, initially, citation of these 'staying behind' publications in other, more cited publications of the group (so the mechanism starts with (within-group) self-citation), and then authors in other groups are stimulated to take notice of these 'staying behind' publications and they decide to cite them. It is obvious that particularly the lower performance groups will benefit from this mechanism.

### 3.3 The concept of hierarchically layered networks of publications and groups

Scale-dependent (cumulative) advantages of research groups are interesting from a viewpoint of network analysis. The basic elements in our study of research groups are publications. Publications act as nodes in a citation network, the links being the citations from other papers, so that the distribution of the number of citations represents the 'in-degree' distribution of the network. These links are (almost) always unidirectional (see Fig. 6): if publication $p_1$ is cited by publication $p_2$, this publication $p_2$ cannot (generally) be cited by $p_1$ (an exceptional case is for instance two publications published together in the same journal issue and citing each other mutually). In our study, this in-degree distribution is given by $\boldsymbol{p}(\boldsymbol{C})$, see Fig. 7a (we refer also to van Raan 2005a for a more detailed discussion of this distribution function).

In the last few years we see a considerable increase of attention from the physics community for publication- and citation-based networks (Redner 1998; Barabási and Albert 1999; Vazquez 2001; Albert & Barabási 2002; Dorogovtsev and Mendes 2002; Mossa *et al* 2002). More recently, the focus of the network community is moving toward clustering phenomena in networks (Klemm and Eguíluz 2002; Ravasz and Barabási 2003) and inter-network relations.

Usually, clustering in a network is defined as a grouping of nodes on the basis of the linkages with which the network is structured. Examples are publication-clusters based on citation or reference linkages (van Raan 2005c), and clusters of collaborating researchers (Newman 2003). This type of clustering is within the network itself. The basic elements (nodes) of a network can, however, also be clustered in another structure, outside their 'own' network, for instance in another network. Research groups can be seen as clusters of publications representing a 'higher order network', based on organizational relations, a 'hierarchical layer' above the network structured on the basis of individual publications. The group as a whole act as a node in the network of research groups. These groups are also connected to each other through citation linkages, but now it is not important which individual publication is 'responsible' for the incoming citations. In the context of this study, ***C*** is the number of incoming links to a group (in-degree of the group), and ***P*** is its size, in terms of total number of publications of the group. In the example of Fig. 6 we find ***P*** = 6 and ***C*** = 8 for group $G_1$. The distribution function of the number of groups over the number of incoming links, ***G***(***C***), is given in Fig. 7b. So, in this model of hierarchically layered networks, we have the in-degree distribution of the basic network (network of the individual publications) represented by ***p***(***C***) (Fig.7a), and the in-degree distribution of the hierarchically next layer (network of the groups) represented by ***G***(***C***) (Fig.7b).

There are striking differences between the higher-order network of groups and the basic publication network. First, in the basic network the nodes (i.e., the publications) do not have a *size* (there is a node, or not), whereas in the group-based network the nodes (i.e., the groups) do have a size, namely, the number of individual publications within a group. Second, the citation-based linkages at the group level are not *unidirectional* anymore (see Fig. 6): if research group $G_1$ is cited by group $G_2$, this group $G_2$ can very well be cited by group $G_1$, within a specific period of time. Third, in the basic network the nodes (i.e., the publications) can only have one link, i.e., publication $p_1$ is cited by publication $p_2$, and this happens only once, whereas research group $G_1$ can be cited many times by group $G_2$.

This third point introduces an interesting analogy with co-author networks. Generally, networks of scientific collaboration based on co-author relations, are structured by linking authors who have at least one common publication. But authors can be linked to other authors by more than just one publication, so authors will in fact have different 'linkage strengths' with other authors, similar to the different citation linkage strengths group $G_1$ has with group $G_2$ (3 citation links) and $G_4$ (1 citation link) in Fig. 6. The analogy goes further: authors also may differ substantially in the size of their own oeuvre, i.e., the number of papers they have, regardless of co-authors. For instance, the link between author $a_1$ (a senior scientist) who has many papers and three links with author $a_2$ who has only these three papers (a PhD student), and just one link with another prolific author $a_3$ (another senior scientist). In the usual scientific collaboration network $a_1$ is simply connected with one link to $a_2$ and with one link to $a_3$, without any indication of the size of $a_1$, $a_2$, and $a_3$. In reality we have a situation comparable with the example in Fig. 6, the relation of the large group $G_1$ (like $a_1$) with the small group $G_2$ (like $a_2$), and $G_1$ with the other large group $G_3$ (like $a_3$).

We stress that this phenomenon is not the same as preferential attachment. Preferential attachment means that highly connected nodes increase their connectivity more than less connected nodes. But in our study of research groups it is not the number of already existing links (***C***, 'external wiring'), but the size ('content'), in terms of number of papers (***P***), that causes a preference, an advantage. We find that, in general, the larger a group (node in the research group network), the larger, in a preferential, i.e., non-linear way (advantage) the 'strength' of the incoming links.

We also observe that the top-10% groups are about an order of magnitude more efficient in 'creating linkages' (***C***) than the bottom-10% groups (see Fig. 3a). Very remarkably, this advantage is not cumulative (exponent 0.72), whereas the bottom-10% groups do have a quite strong cumulative advantage (exponent 1.44), so that for a specific size (***P***), top-10% and bottom-10% have almost the same strength (***C***) at the same size.

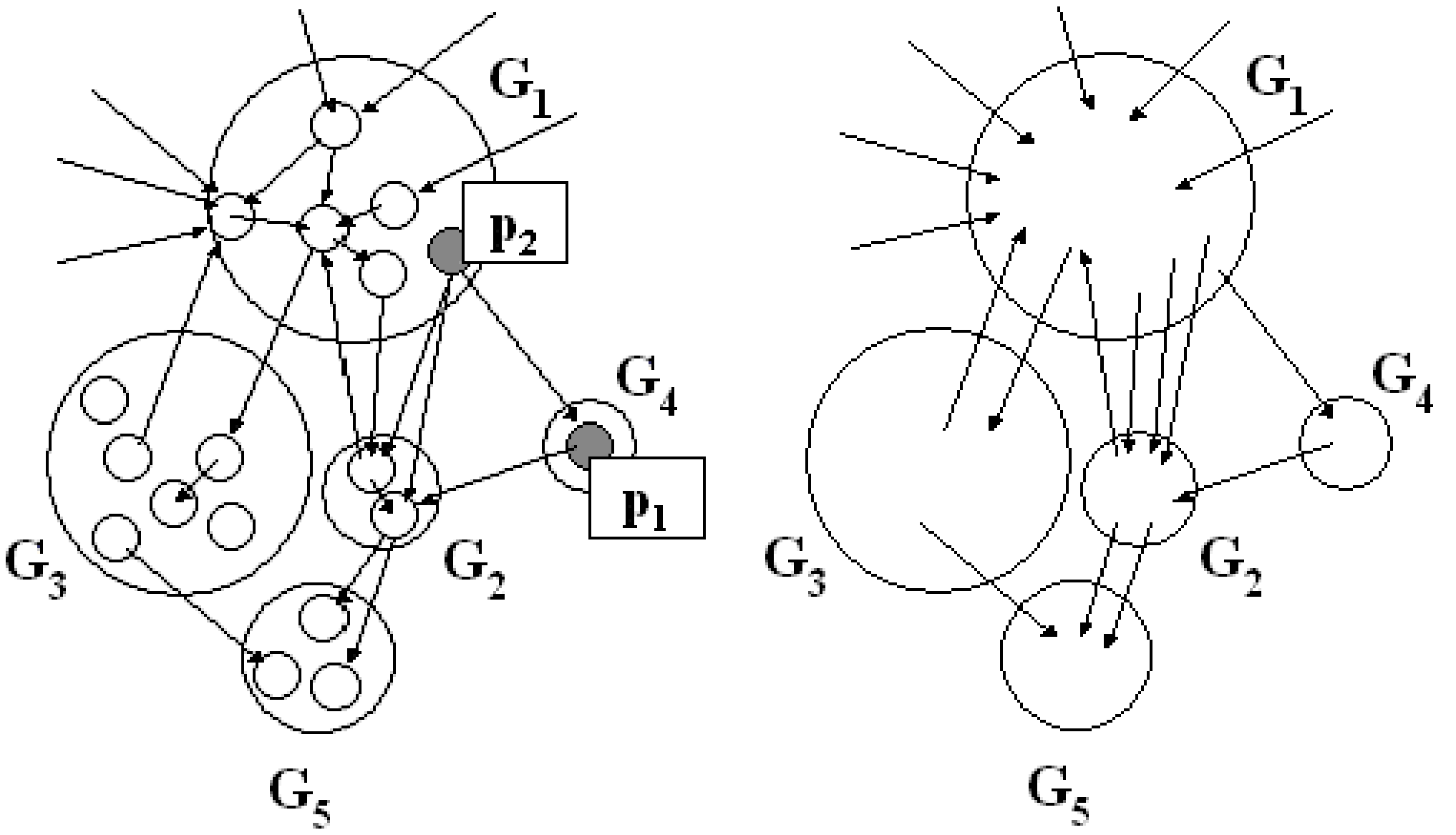

***Fig. 6,*** *left side: network structure (example) of individual publications with incoming links (citations to a publication, unidirectional), the in-degree distribution is given by* ***p*(*C*)***, see Fig. 7a; right side: same network structure now at the level of research groups (bi-directional links allowed), the in-degree distribution is given by* ***G*(*C*)***, see Fig. 7b.*

As our criterion concerning top-performance or lower performance is based on the field-normalized performance indicator ***CPP/FCSm***, we hypothesize that in networks terms this indicator represents the *fitness* of a group as a node in the group-network. It brings a group in a better position to acquire additional links on the basis of mere *size* (an 'internal' parameter), and not on the basis of already existing linkages to a node (an 'external' parameter). This latter mechanism, preferential attachment, is based on the idea that other nodes, for instance newcomers, feel 'directly' the attractiveness of a node and therefore also want to have a link with the already attractive node. In our explanation, size of the node is crucial, with a simple 'advantage-making' mechanism as explained at the end of Section 3.2.

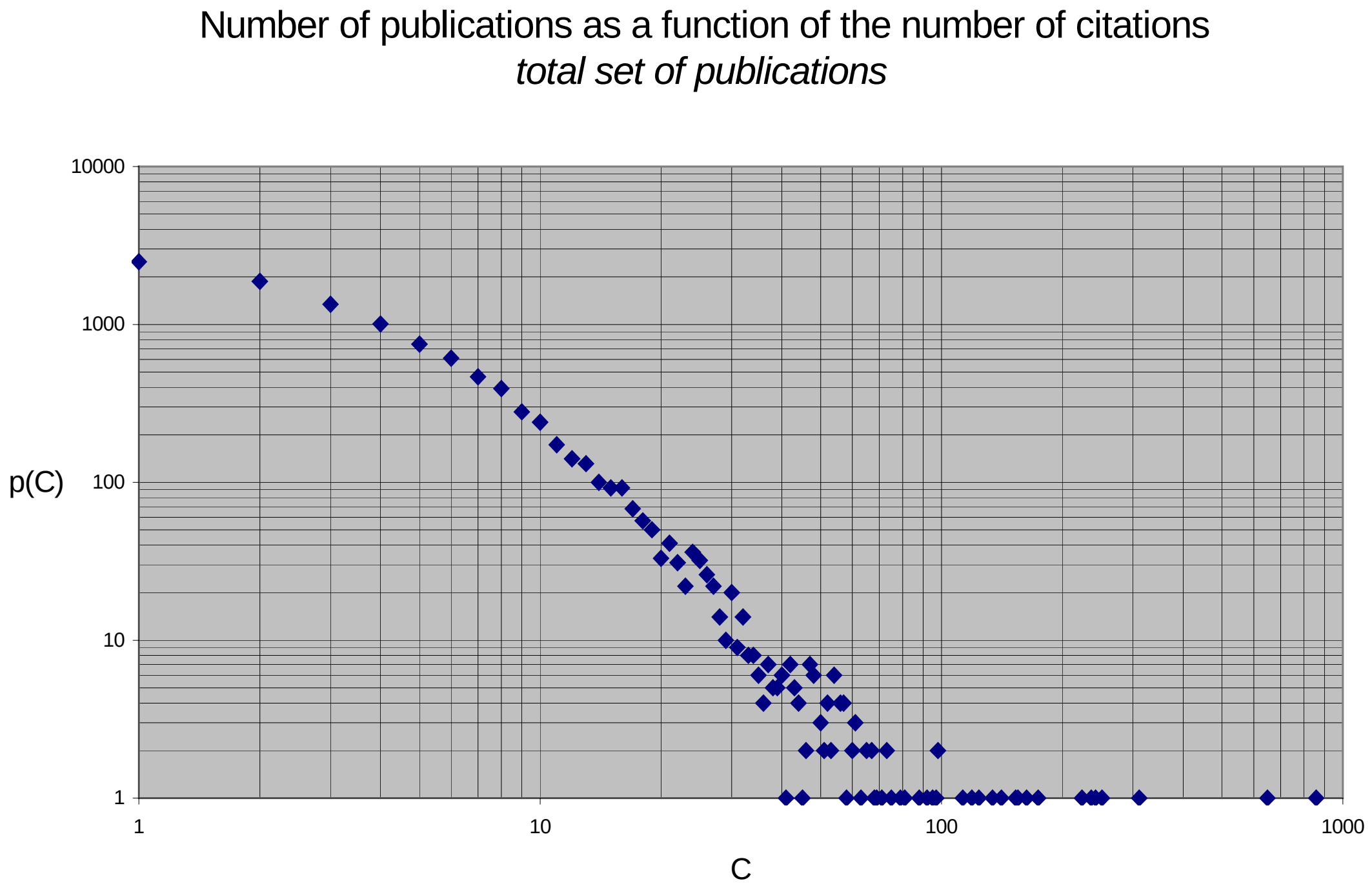

***Fig. 7a**: Distribution function **p**(**C**): number of publications as a function of number of citations for the total set of around 14,000 chemistry publications.*

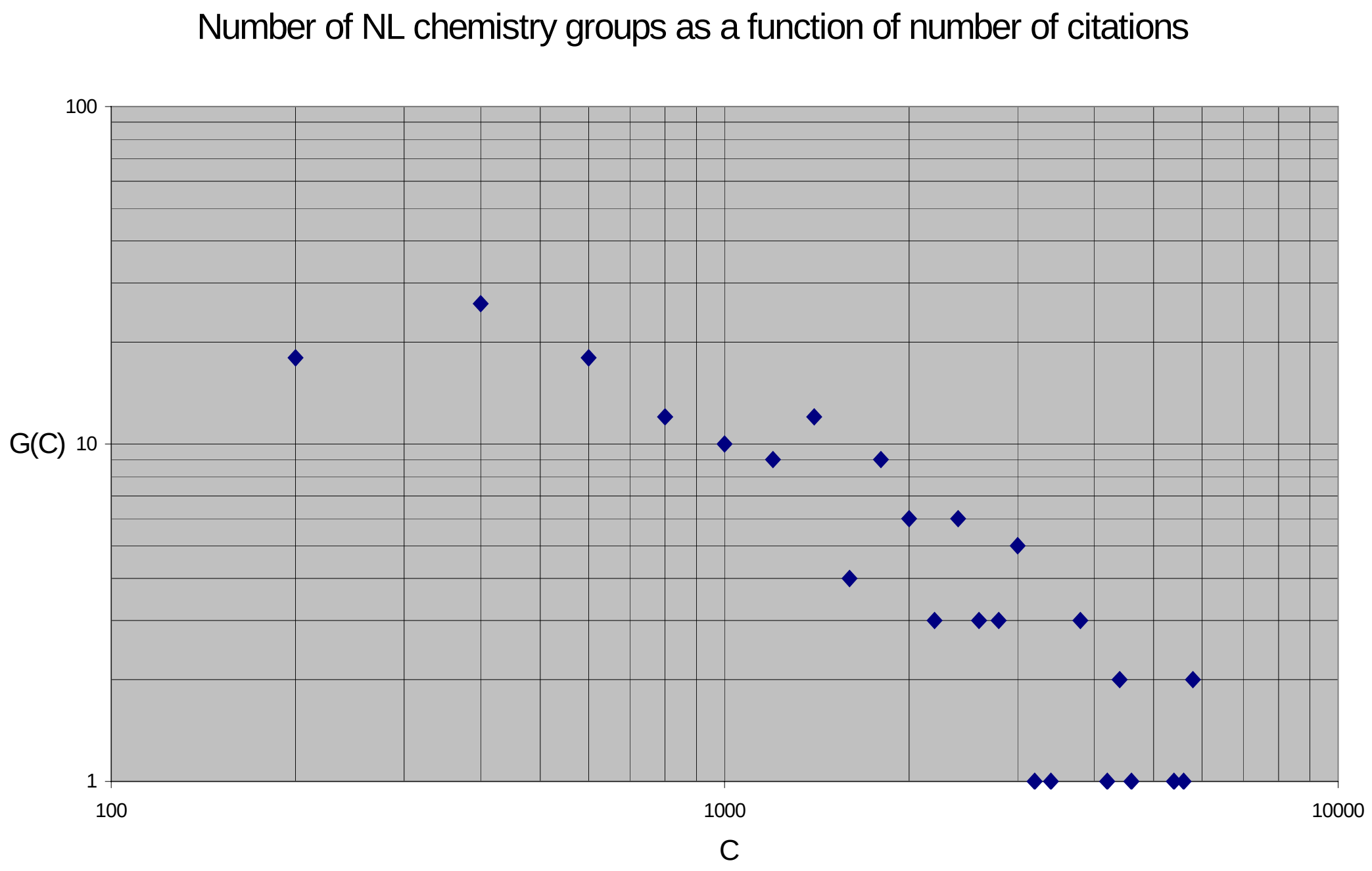

***Fig. 7b**: Distribution function **G**(**C**): number of groups as a function of number of citations per group, for the total set of the 157 chemistry research groups; groups are binned in classes of Δ **C** = 200.*

### 3.4 Scale-dependence of 'fitness'

We find no or hardly any significant dependence of our 'fitness' indicator ***CPP/FCSm*** with size in terms of ***P***, as shown in Figs. 8a and b for the top-50% and the bottom-50% (of ***CPP/FCSm*** scores) of the chemistry and the medical research groups, respectively. Further analysis with the top- and bottom 10% and the top- and bottom 20% reveals that only for the bottom-10% and bottom-20% of the medical research groups there is a reasonably significant ($R^2$ = 0.55 and 0.51, respectively) correlation with ***P***. This indicates that for the lower performance medical research groups there is some positive correlation with size, but this is certainly not a cumulative advantage as the exponents of the correlation are 0.14 and 0.15, respectively, more or less similar to the situation as in the case of the bottom-50% of the medical research groups, as shown in Fig.8b.

We notice that for very large size ***P*** the value of ***CPP/FCSm*** must go asymptotically to 1, because the largest possible ***P*** would be all publications worldwide in a specific field (or combination of fields), which means that by definition the field-normalized indicator values has to be 1. Remarkably, in our observations an onset to this asymptotic behaviour is only slightly visible for the top groups in chemistry.

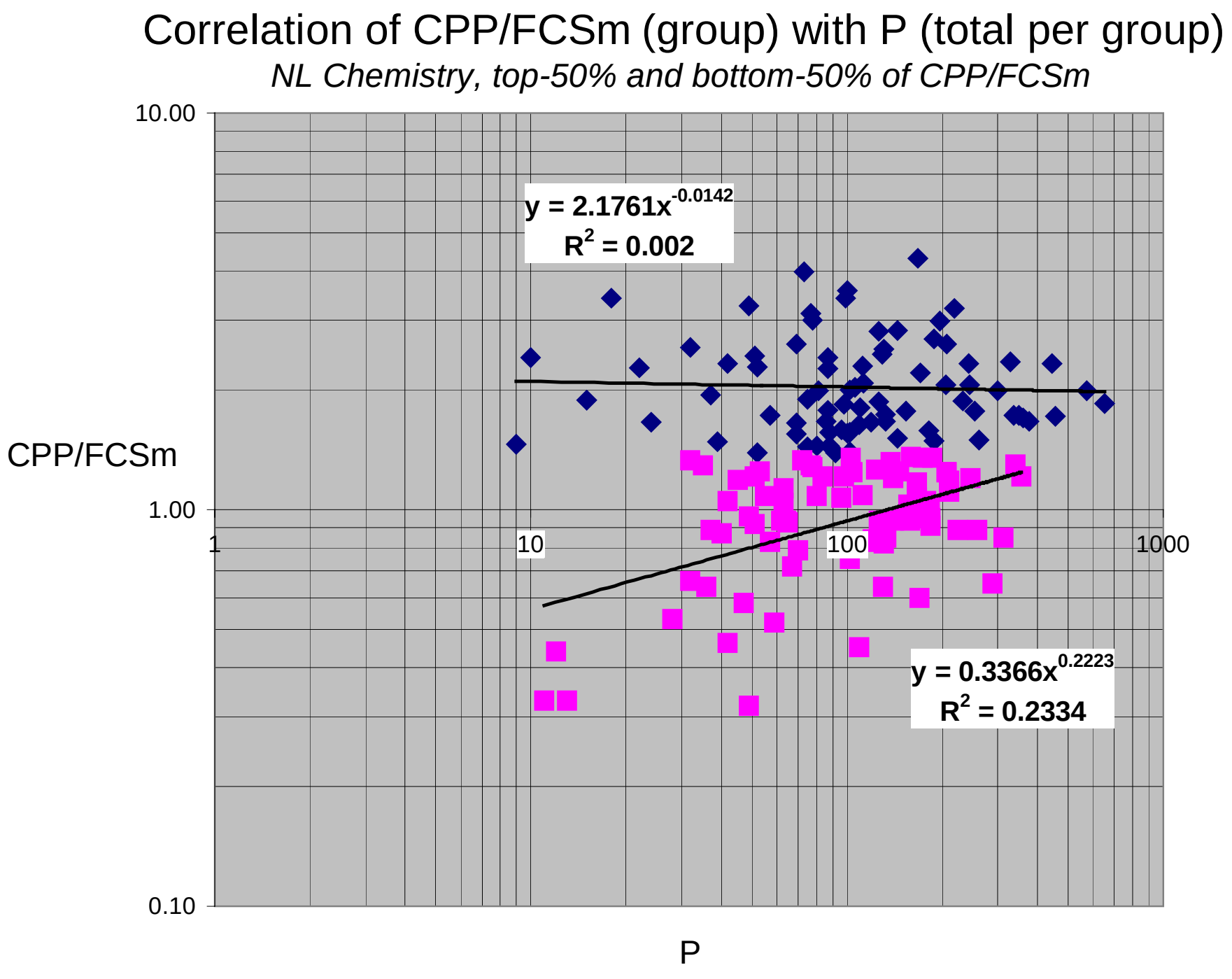

***Fig. 8a:*** *Correlation of* ***CPP/FCSm*** *with the number of publications (**P**), for the top-50% (of* ***CPP/FCSm****) chemistry groups (indicated with diamonds), and for the bottom-50% chemistry groups (indicated with squares).*

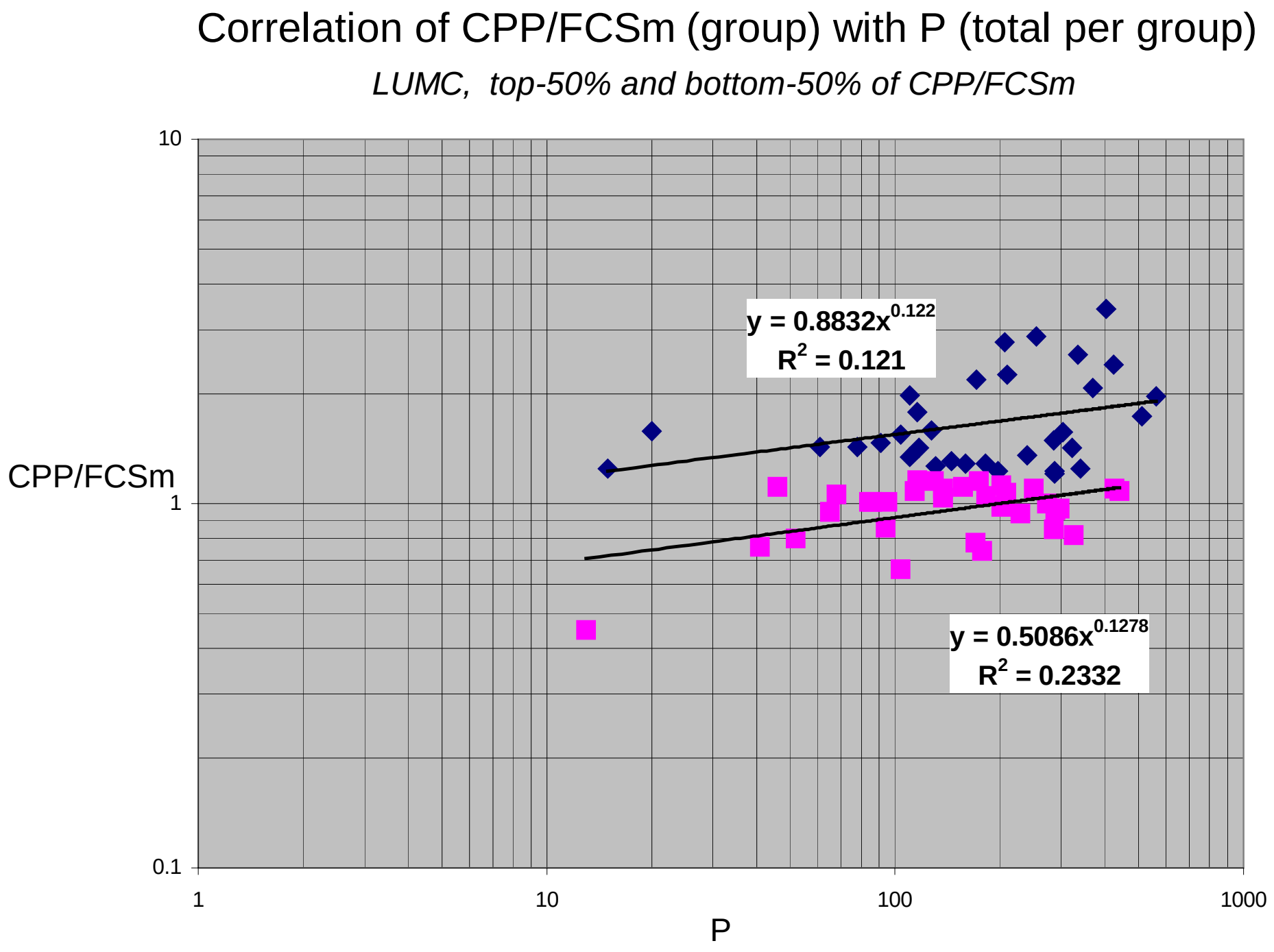

***Fig. 8b:*** *Correlation of* ***CPP/FCSm*** *with the number of publications (****P****), for the top-50% (of* ***CPP/FCSm****) medical research groups (indicated with diamonds), and for the bottom-50% medical research groups (indicated with squares)*.

## 3.5 Publication citedness and journal impact level

Seglen (1994) reported on the poor correlation between the impact of publications and journal impact at *the level of individual publications*. However, grouping publications in classes of journal impact yielded a high correlation between publication citedness and journal impact. But this higher aggregation is determined by journal impact classes, and not by a 'natural' higher aggregation level such as a research group.

We find a quite significant correlation between the average number of citations per publication ('publication citedness', given by the indicator ***CPP***) of research groups, and the average journal impact of these groups (given by the journal impact indicator ***JCSm***). The results are shown in Figs. 9a and b for the entire sets of all chemistry and medical research groups, respectively.

By dividing the authors into a highly cited group and a less cited group, Seglen concluded that the highly cited authors tend to publish somewhat more in journals with a higher impact than the less cited authors, but this difference is insufficient to explain the difference in impact between the two groups. Highly cited authors are, according to Seglen, in all journal impact classes, on average, more successful. Thus, we applied again the distinction between top-performers and lower-performance groups as used throughout this paper in order to find performance-dominated aspects in the relation between publication citedness and journal impact level.

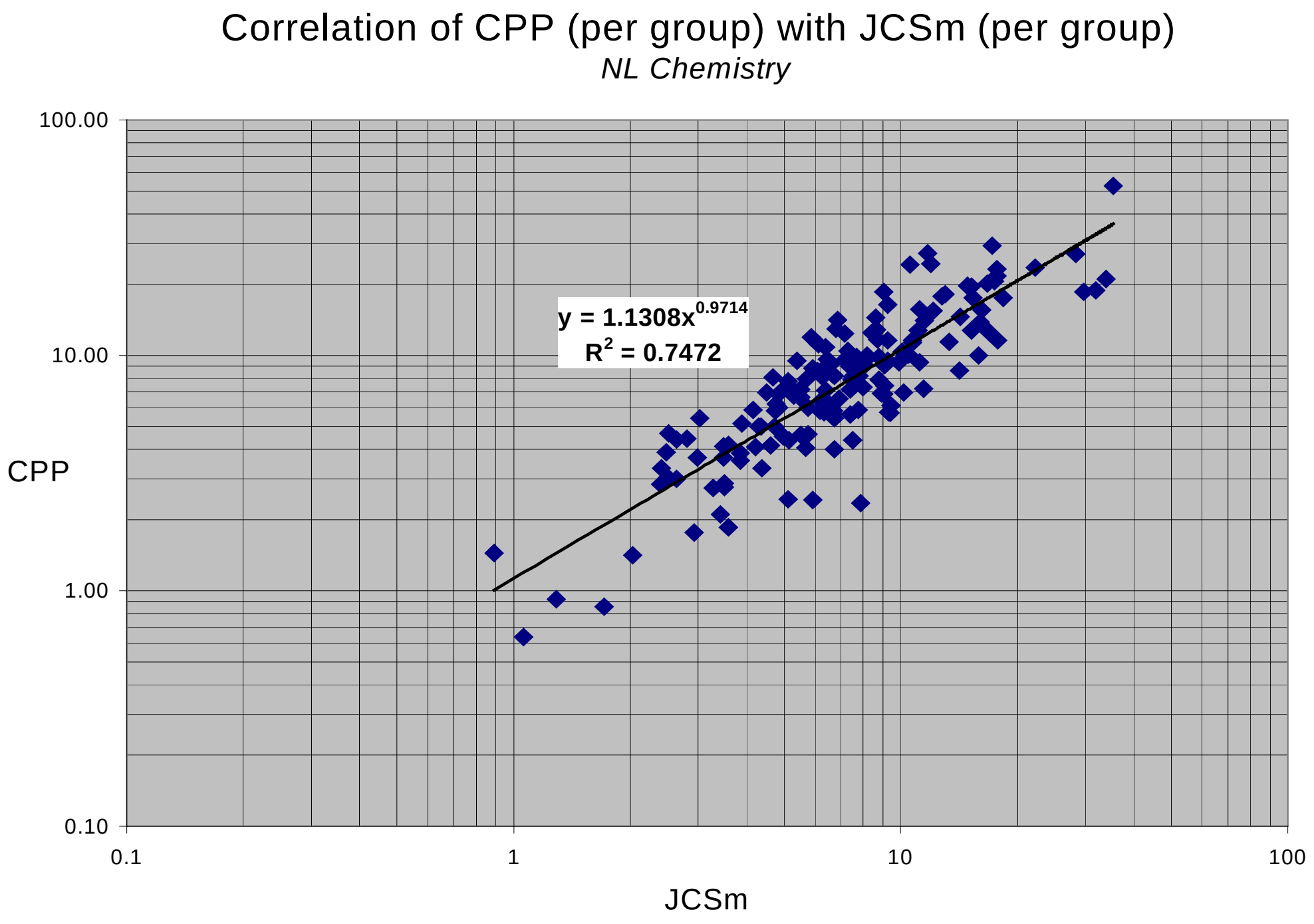

***Fig. 9a:*** *Correlation of* ***CPP*** *with the* ***JCSm*** *values for all chemistry groups.*

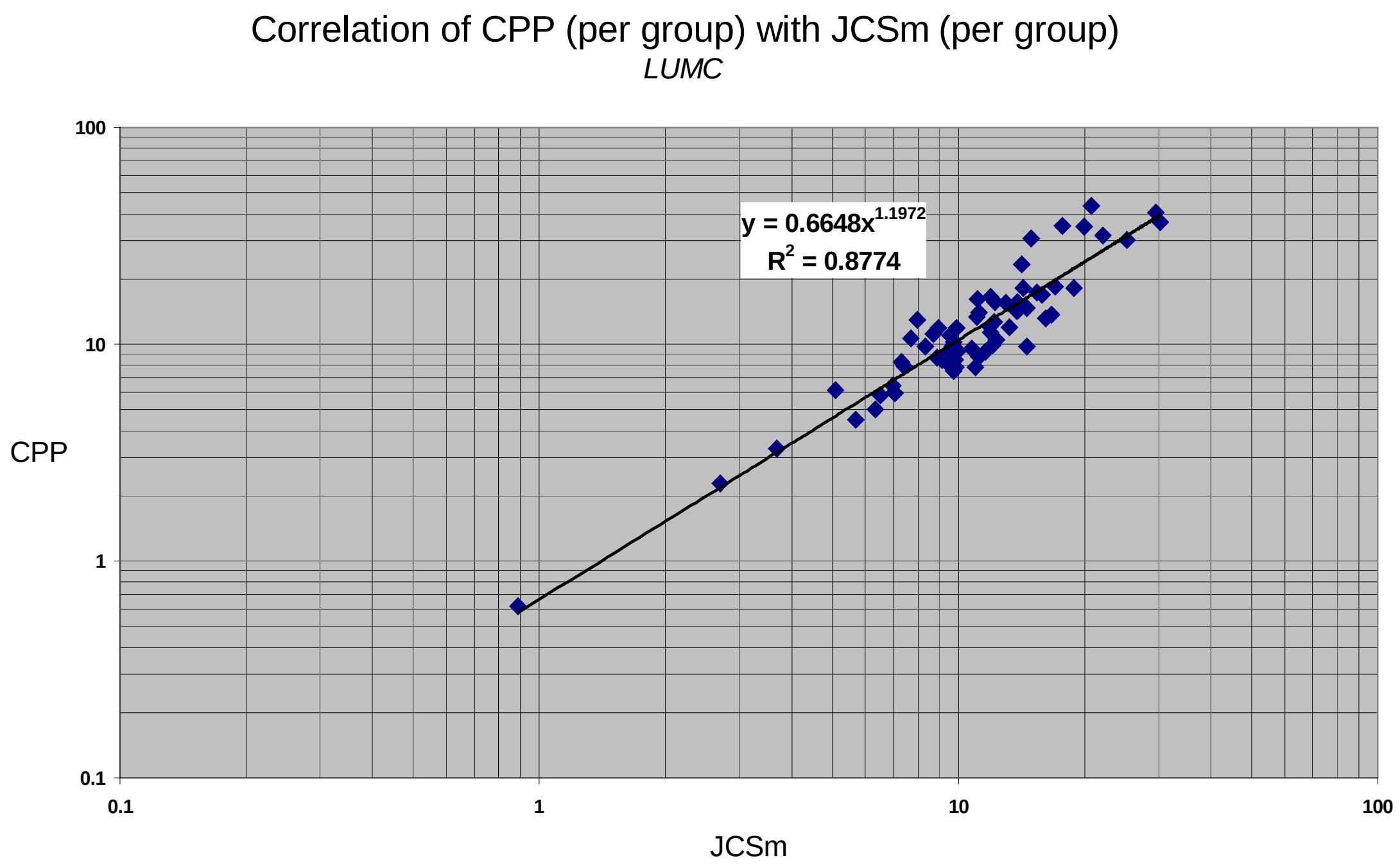

***Fig. 9b:*** *Correlation of* ***CPP*** *with the* ***JCSm*** *values for all medical research groups.*

Following the same procedure as in Section 3.1, we first created within the entire set of chemistry groups two subsets on the basis of the quality judgement by peers. One subset with 39 'top-performance' groups, these groups received the highest judgment 'excellent' (***Q*** = 5), and another subset with 30 lower performance groups, these groups that received the lowest judgment 'satisfactory' (***Q*** = 3). The results are given in Fig. 10.

We clearly observe the differences *and similarities* between the two subsets. Both the 'excellent' as well as the 'satisfactory' groups generally have more total citations per publication (***CPP***) as a function of journal impact (***JCSm***). Clearly, the excellent groups generally have higher ***CPP*** values. Remarkably, the excellent as well as the satisfactory groups are more or less in the same range of journal impact values. Thus, these observations nicely confirm Seglen's findings as discussed above. Indeed, top-performance groups are, on average, more successful in the entire range of journal impact. In other words, they perform better in the lower-impact journals as well as in the higher impact journals. Next, we also notice that there is no 'cumulative advantage', i.e., the power law exponent of the correlation function is about (excellent groups) or below (satisfactory groups) value 1.

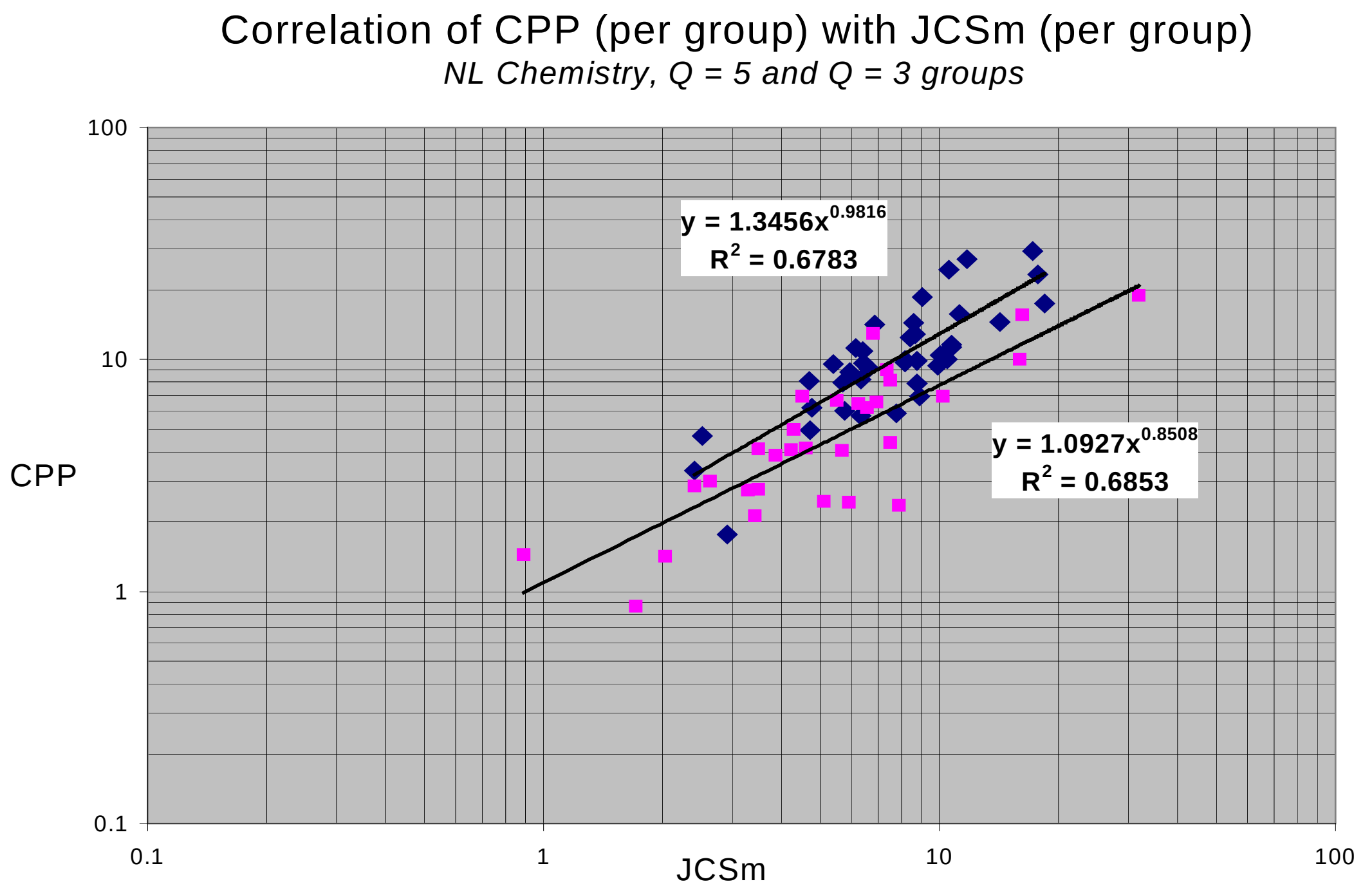

***Fig. 10:*** *Correlation of* ***CPP*** *with the* ***JCSm*** *values for the top-performance chemistry groups (Q=5, indicated with diamonds), and for the lower performance groups (Q=3, indicated with squares).*

We now carry out the same analysis on the basis of our own research performance (field-normalized impact) indicator ***CPP/FCSm***. We created within the entire set of chemistry research groups and medical research groups the following subsets: groups belonging to the top-10%, top-20%, and top-50%, as well as to the bottom-10%, bottom-20%, and bottom-50% of the ***CPP/FCSm*** distribution. We present the results of the correlation measurement in Figs. 11a and 11b (top-10% compared to bottom-10%, chemistry and medical groups, respectively), 11c and 11d (top-20% compared to bottom-20%, chemistry and medical groups, respectively), and 11e and 11f (top-50% compared to bottom-50%, chemistry and medical groups, respectively).

We observe the same phenomena for the chemistry research groups as found in Fig. 10. However, the top-10% of the ***CPP/FCSm*** distribution is clearly somewhat more 'exclusive' compared to the groups with the peer judgement 'excellent': these top-groups do have a (slight) preference for the higher-impact journals (Fig.11a). For the medical groups this phenomenon is even more pronounced (Fig. 11b): the top-10% (and also top-20%, Fig. 11d) medical groups appear to focus heavily on the high-impact journals!

In the case of the chemistry groups, the ratio of the correlation coefficients between ***CPP*** and ***JCSm*** provides a quantitative measure of the extent to which top groups have a higher 'citedness' as compared to lower performance groups. For top-10% and bottom-10% the ratio of the correlation coefficients is 3.45, for top-20% and bottom-20% it is 3.07, and for top-50% and bottom-50% we find 1.87. This means that the chemistry top-groups perform (in terms of ***CPP***, citations per publication) with a factor of about 2 to 3.5 better than the bottom-groups *in the same journals*. Also this finding is in agreement with Seglen's work, he finds a factor between 1.5 and 3.5.

Finally, we analyse the exponents of the correlation functions presented in Figs. 11a-f and present the results in Table 4.

***Table 4***: *Power law exponent γ of the correlation of **CPP** with **JCSm** for the two sets of groups, in the indicated modalities. The differences in γ between the set of chemistry research groups and the set of medical research groups is given by Δγ(M,C); the difference between the top and bottom modalities (see text) by Δγ(b,t). The value between parentheses has a low significance, hence no differences as indicated above are calculated.*

| | **Chemistry Groups** | **Medical Groups** | ***Δγ**(M,C)* |
|---|---|---|---|
| **top 10%** | 0.91 | (0.59) | |
| **bottom 10%** | 0.94 | 1.06 | *0.12* |
| ***Δγ**(b,t)* | *0.03* | | |
| | | | |
| **top 20%** | 0.90 | 0.97 | *0.07* |
| **bottom 20%** | 1.03 | 1.05 | *0.02* |
| ***Δγ**(b,t)* | *0.13* | *0.08* | |
| | | | |
| **top 50%** | 0.90 | 1.17 | *0.27* |
| **bottom 50%** | 0.96 | 1.05 | *0.09* |
| ***Δγ**(b,t)* | *0.06* | *-0.12* | |

In general, we see correlation function exponents close to 1, which means that the number of citations per publication is about a linear function of journal impact. By randomly removing 10 groups in the set of chemistry research groups and recalculating the correlation functions, we estimate that the uncertainty in the power law exponents is about ±0.04. Therefore we conclude that only in the case of the medical research groups there might be a slight 'cumulative' advantage of ***CPP*** with ***JCSm***, particularly for the top-50%.

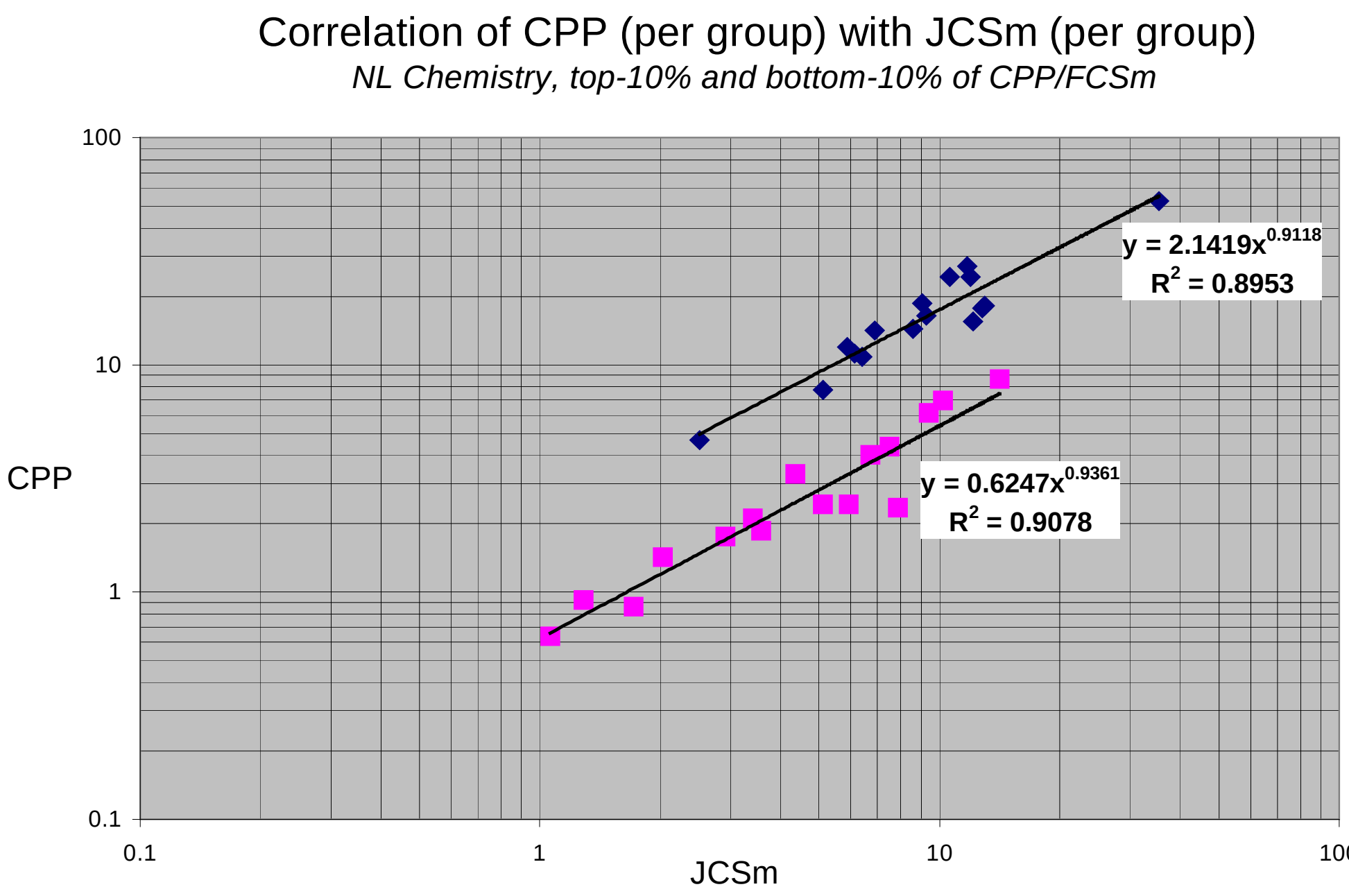

***Fig. 11a:*** *Correlation of the number of citations (**C**) received per chemistry research group with the number of publications (**P**), for the top-10% (of **CPP/FCSm**) groups (indicated with diamonds), and for the bottom-10% groups (indicated with squares).*

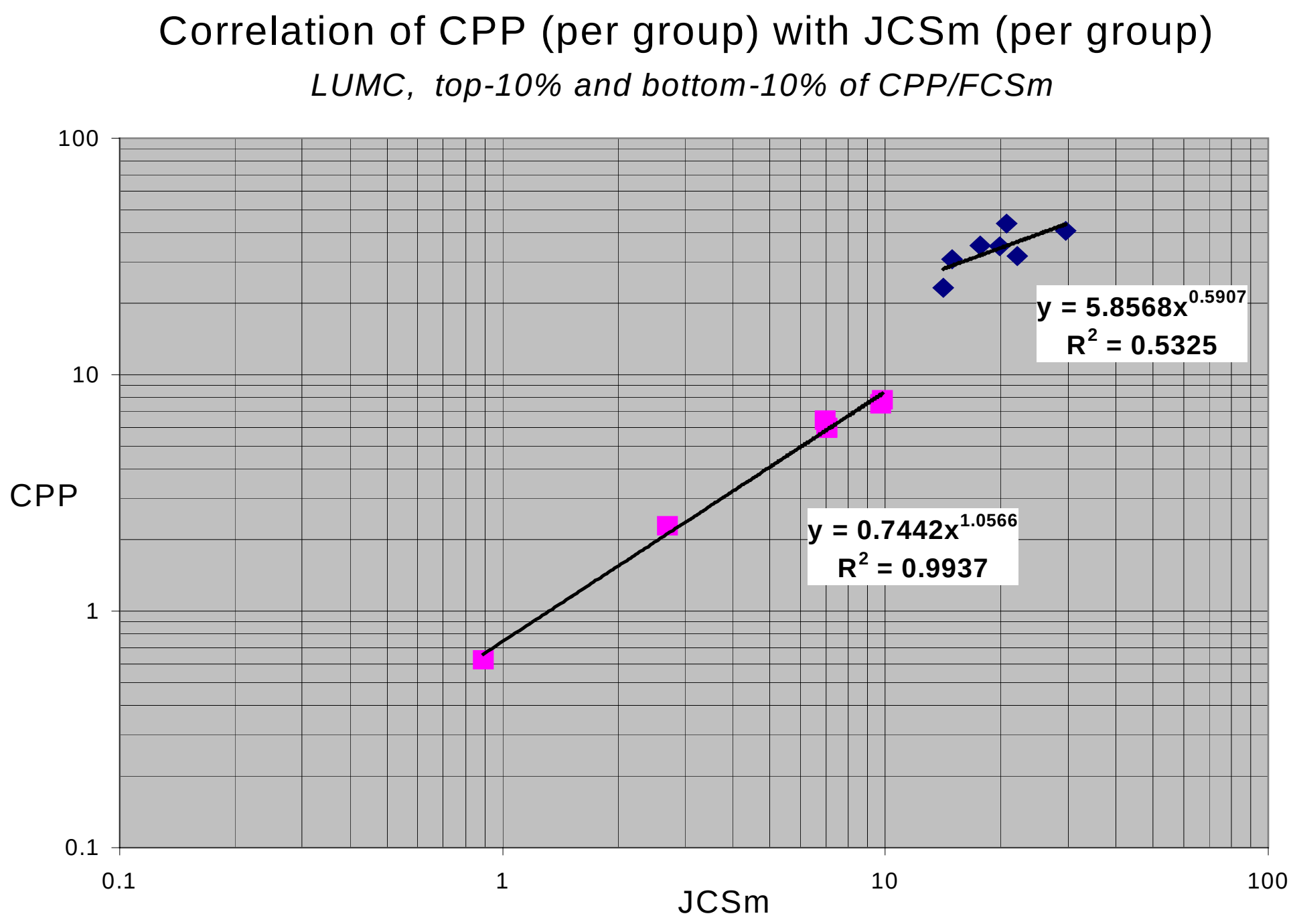

***Fig. 11b:*** *Correlation of the number of citations (**C**) received per medical research group with the number of publications (**P**), for the top-10% (of **CPP/FCSm**) groups (indicated with diamonds), and for the bottom-10% groups (indicated with squares).*

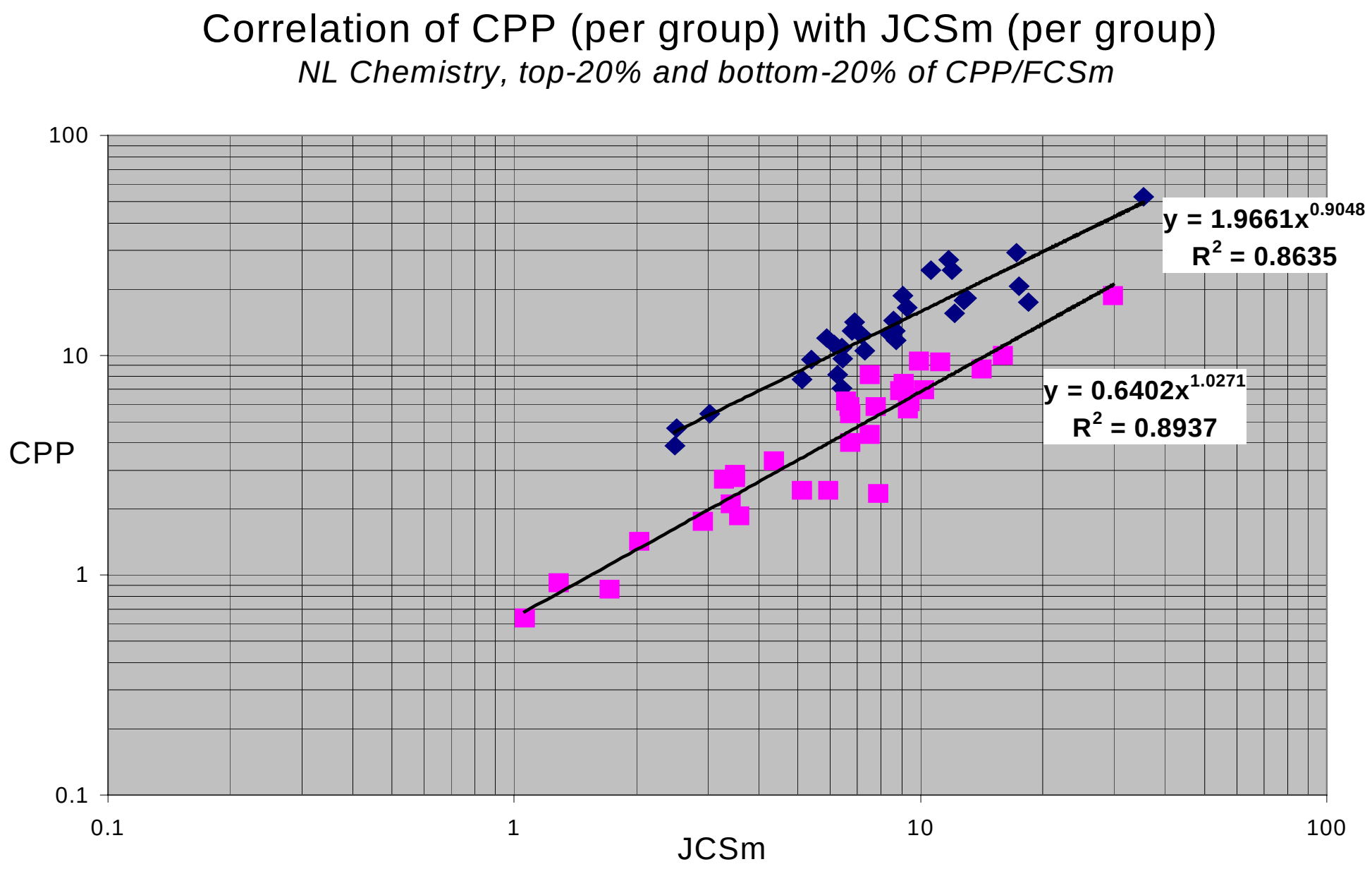

***Fig. 11c:*** *Correlation of the number of citations (**C**) received per chemistry research group with the number of publications (**P**), for the top-20% (of **CPP/FCSm**) groups (indicated with diamonds), and for the bottom-20% groups (indicated with squares).*

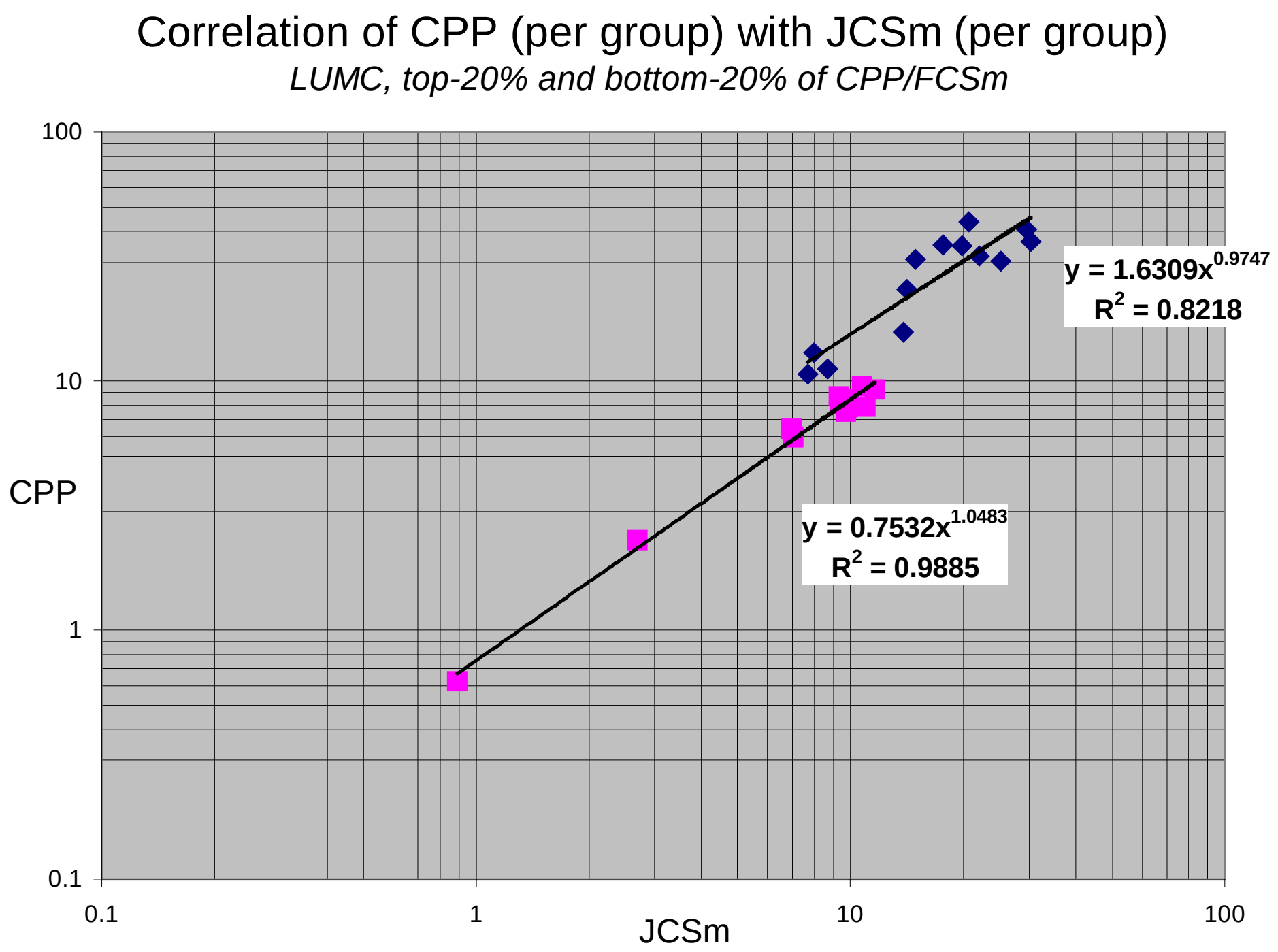

***Fig. 11d:*** *Correlation of the number of citations (**C**) received per medical research group with the number of publications (**P**), for the top-20% (of **CPP/FCSm**) groups (indicated with diamonds), and for the bottom-20% groups (indicated with squares).*

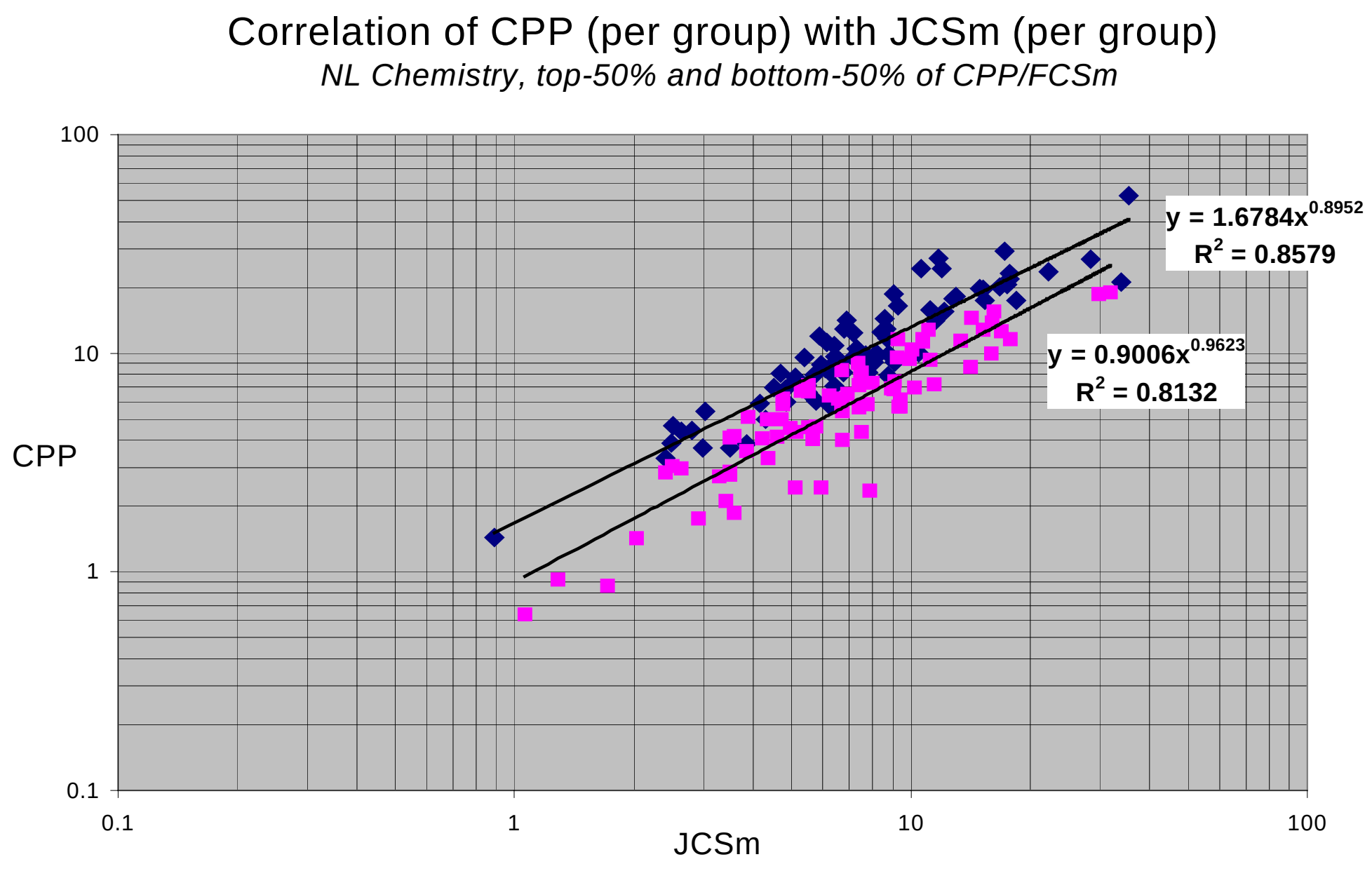

***Fig. 11e:*** *Correlation of the number of citations (**C**) received per chemistry research group with the number of publications (**P**), for the top-50% (of **CPP/FCSm**) groups (indicated with diamonds), and for the bottom-50% groups (indicated with squares).*

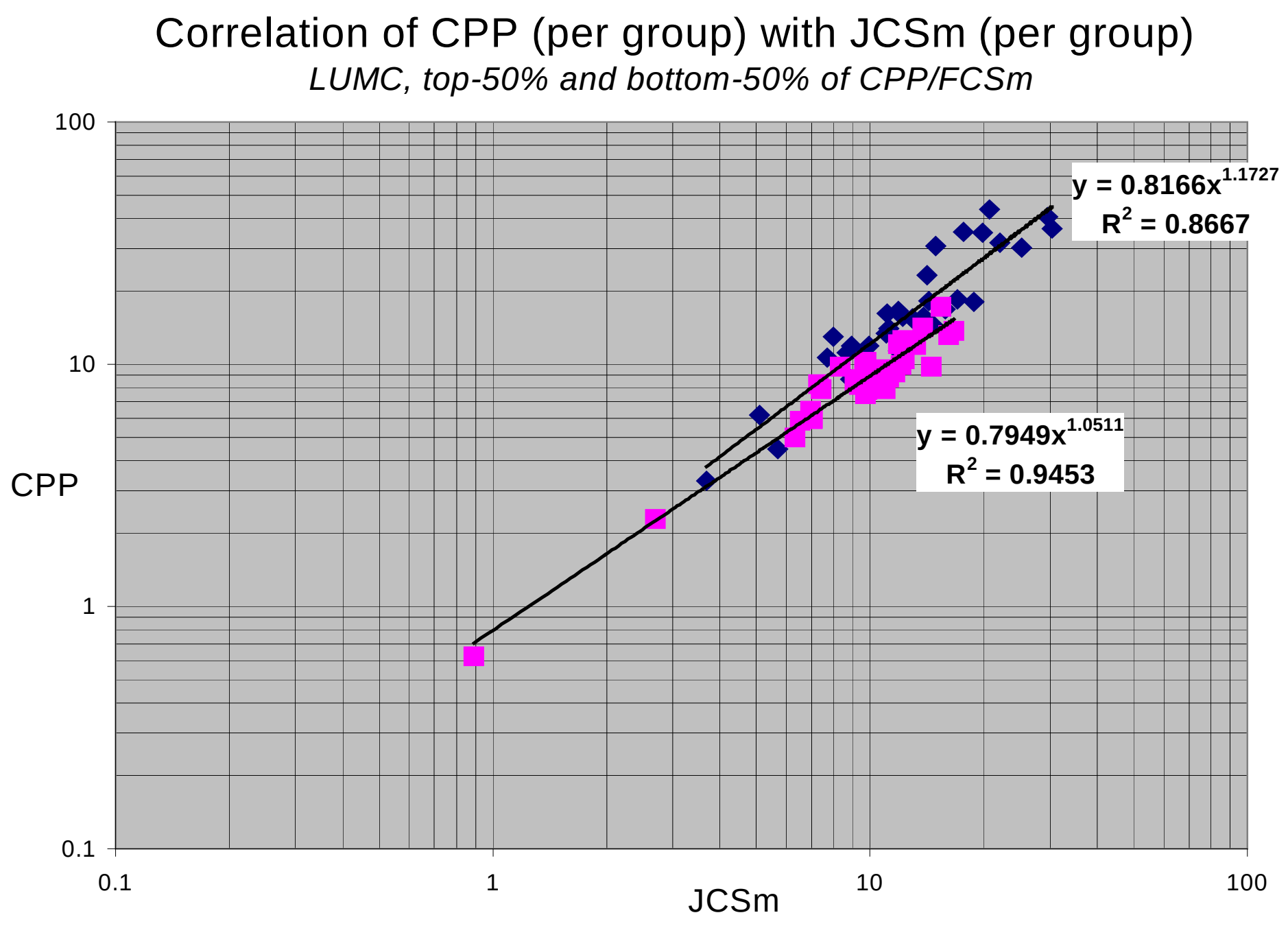

***Fig. 11f:*** *Correlation of the number of citations (**C**) received per medical research group with the number of publications (**P**), for the top-50% (of **CPP/FCSm**) groups (indicated with diamonds), and for the bottom-50% groups (indicated with squares).*

Finally, we analysed the correlation between the number of not-cited publications (***Pnc***) of a group and its average journal impact level (***JCSm***). The results for the medical research groups are shown in Fig. 12a. We see a quite significant correlation between these two parameters. Given the strong correlation between ***CPP*** and ***JCSm*** (see Fig. 9b), we can expect also a significant correlation between ***Pnc*** and ***CPP***, as confirmed by Fig. 12b. We observe that the higher the mean number of citations in a group, the lower the number of not-cited publications in a group. In other words: groups that are cited more per paper also have more cited papers. These findings underline the generally good correlation (at the group level, not at the individual publication level!) between the mean 'citedness' of publications in a group, and its mean journal impact.

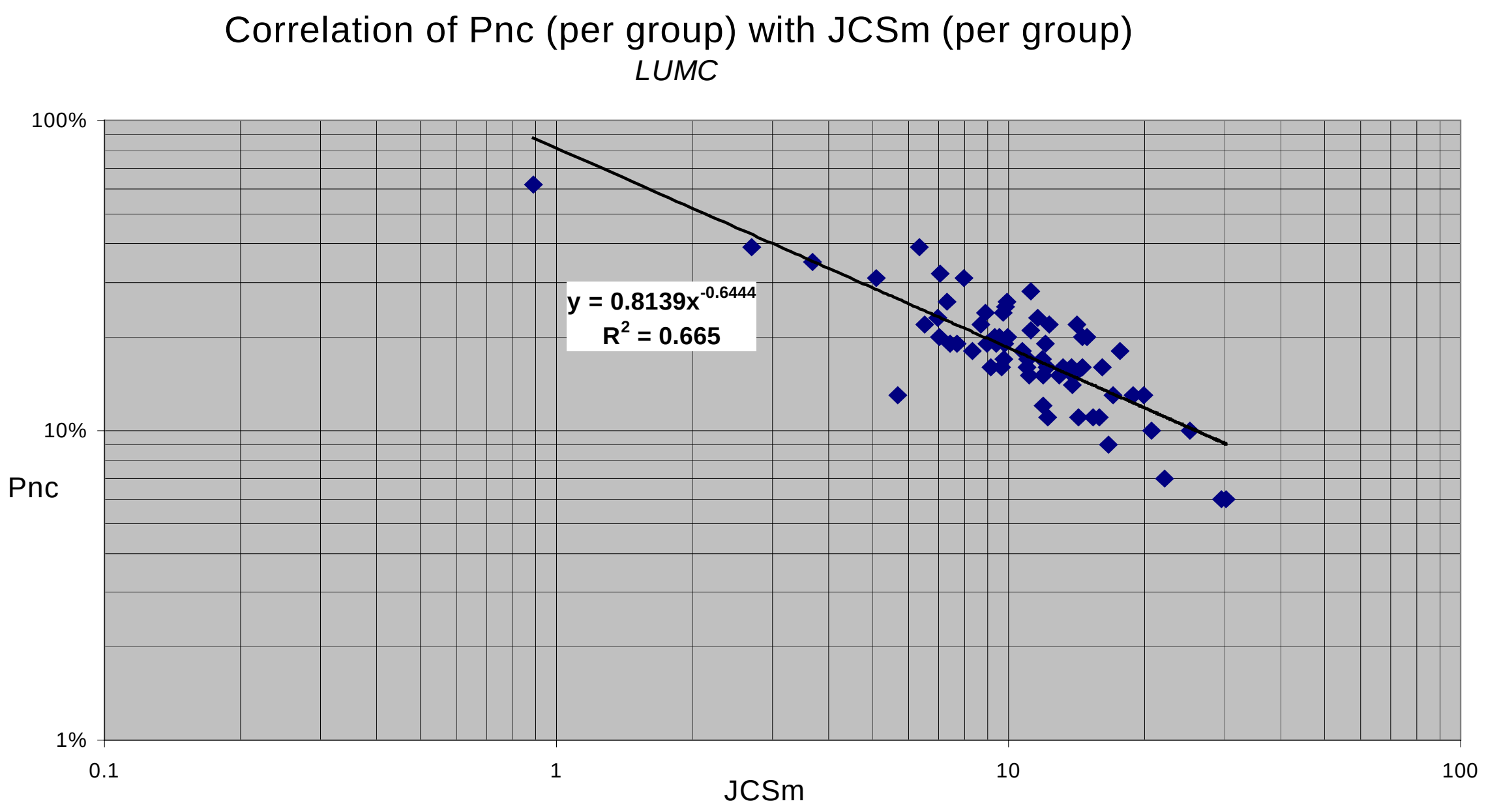

***Fig. 12a:*** *Correlation of the number of not cited publications (**Pnc**) of medical research groups with the mean journal impact (**JCSm**) of a group.*

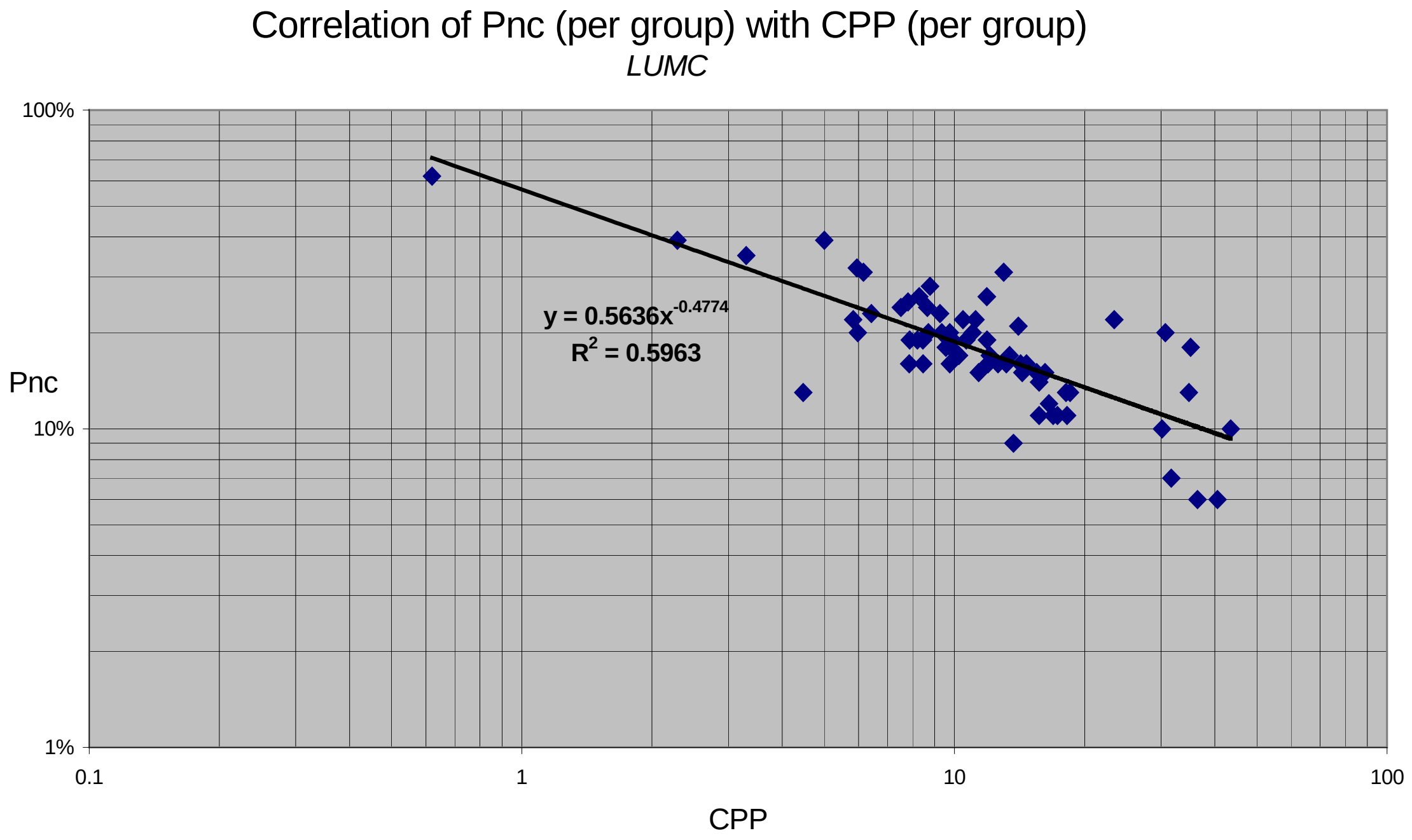

***Fig. 12b:** Correlation of the number of not cited publications (**Pnc**) of medical research groups with the mean number of citations per publications (**CPP**) of a group.*

## 4. Summary of the main findings and concluding remarks

We studied performance-related statistical properties of bibliometric characteristics of two sets of research groups: 157 chemistry groups and 65 medical research groups, covering a period of at least ten years. Our main observations are as follows.

First, we find a size-dependent cumulative advantage for the total impact of research groups in terms of total number of citations. Quite remarkably, the lower-performance groups show a stronger cumulative advantage.

Second, we find that, regardless of performance, larger groups have less not-cited publications. We introduce a simple model in which size is advantageous in an 'internal promotion mechanism' to get more publications cited. Thus, in this model size is a distinctive parameter and it acts as a bridge between the macro picture (characteristics of the entire set of groups) and the micro picture (characteristics within a group).

Third, by distinguishing again between top- and lower-performance groups, we discovered that particularly for the lower-performance groups the size-effect mentioned in our second observation is effective, and that the fraction of not-cited publications is decreasing with size. This is in line with a further observation that particularly the lower performance groups tend to have a higher field-normalized impact indicator value the larger these groups are.

Fourth, we find a quite significant correlation between the average number of citations per publication ('publication citedness', given by the indicator ***CPP***) of research groups, and the average journal impact of these groups (given by the journal impact indicator ***JCSm***). Top-performance groups are, on average, more successful in the entire range of journal impact, with a factor of about 2 to 3.5. In other words, top-groups perform better in the lower-impact journals as well as in the higher impact journals. There is no clear evidence of cumulative advantage for the citedness of publications with journal impact. Only in the case of the medical research groups there might be a slight cumulative advantage, particularly for the top-groups.

An important element of this study is that we make a first attempt to fit our findings into a concept of 'hierarchically layered' networks. In this concept, the network of research groups constitutes a layer of one hierarchical step higher than the basic network of publications connected by citations. The cumulative size-advantage of citations received by a group looks like preferential attachment in which highly connected nodes increase their connectivity faster than less connected nodes. But in our study of research groups it is not the number of already existing links (***C***, 'external wiring'), but the size ('content'), in terms of number of papers (***P***), that causes a preference, an advantage. We find that, in general, the larger a group (node in the research group network), the larger, in a preferential, i.e., non-linear way (advantage) the 'strength' of the incoming links. Moreover, we find that top-performance groups are about an order of magnitude more efficient in 'creating linkages' (***C***) than the lower performance groups.

As our criterion concerning top-performance or lower performance is based on the field-normalized performance indicator ***CPP/FCSm***, we hypothesize that in network terms this indicator represents the *fitness* of a group as a node in the group-network. It brings a group in a better position to acquire additional links on the basis of mere *size* (an 'internal' parameter), and not on the basis of already existing linkages to a node (an 'external' parameter). Thus, size of the node is crucial, with a simple 'advantage-making' mechanism as mentioned in our second observation.

We 'translate' in this study typical bibliometric properties into network-related properties: ***C***, the number of citations to a group, or total impact, is the 'external wiring' of the group as node in the network; ***P***, the number of publications is the size of the group as network node; ***CPP***, the size-normalized impact; ***CPP/FCSm***, the field-normalized impact, is the fitness of a group as a node in a network structure. The field-based impact ***FCSm*** can be seen as a general local property for a family of groups in the network. How does the journal impact indicator ***JCSm*** fit into this picture? We think ***JCSm*** can be conceived of as a group- (as thus node-) internal characteristic such as a basic facility that is available to 'make the most of it'. As an analogy in a social context one could think of education level (van Raan 2005a): a higher level offers the possibility to reach a higher income, but this is not an automatism, and with a relatively low education level one has still has a chance for a high income (in network terms: external wiring, or incoming links).

Next to the intriguing differences between top-performance and lower performance groups, we also find differences between the two sets of research groups. The chemistry groups are from ten different universities, they have grown more or less 'naturally', and they are not subject to one specific research policy strategy as all these ten universities have their own priorities. The medical groups, however, are all within one large institution. Although they also can be considered as having a 'natural' basis as a research group around one or two full professors, these groups are at the same time influenced by the policy of the LUMC as a whole. Close mutual collaboration and the availability of the best people and facilities of a wide range of groups in the same location may enhance performance. Currently we are extending this study to more large sets of research groups in different scientific fields in order to investigate whether the differences found in this study between the chemical and the medical research groups are indeed due to research management related aspects, or that discipline-related aspect play a dominant role.

***Acknowledgements***

The author would like to thank his CWTS colleague Thed van Leeuwen for the data collection, data analysis and calculation of the bibliometric indicators for the two sets of research groups.

version 070405